\documentclass[aps,prb,twocolumn,floats,amsmath,amssymb,,superscriptaddress]{revtex4-2}
\usepackage[utf8]{inputenc}   
\usepackage[T1]{fontenc}      
\usepackage[english]{babel}   
\usepackage{lmodern}          
\usepackage{csquotes}         
\usepackage{amsmath}          
\usepackage{mathtools}        
\usepackage{amstext}          
\usepackage{amssymb}          
\usepackage{bbm}              
\usepackage{subdepth}         
\usepackage{xcolor}           
\usepackage{placeins}
\usepackage{braket}           
\usepackage{graphicx}         
\usepackage[pass]{geometry}   
\usepackage[pdftex]{hyperref} 
\usepackage[capitalize]{cleveref}

\newcommand{\cee}{\hat{c}^{\vphantom{\dagger}}}

\newcommand{\cdag}{\hat{c}^\dagger}

\newcommand{\wick}{T_\tau}

\renewcommand{\i}{\mathrm{i}}

\newcommand*{\dif}{\mathop{}\!\mathrm{d}}

\newcommand{\vect}[1]{\mathbf{#1}}

\newcommand{\abinitio}{\textit{ab initio}}
\newcommand{\dga}{D$\Gamma$A}
\newcommand{\bfdga}{\textbf{D$\mathbf{\Gamma}$A}}
\newcommand{\itdga}{\textit{D$\mathit{\Gamma}$\!A}}
\newcommand{\scdga}{sc-D$\Gamma$A}
\newcommand{\osdga}{1-D$\Gamma$A}
\newcommand{\abinitiodga}{{\em ab initio} D$\Gamma$A}

\makeatletter
\newcommand{\thefontsize}{\f@size pt\par}
\makeatother

\allowdisplaybreaks[1]    
\hypersetup{
  pdfauthor = {Josef Kaufmann, Christian Eckhardt, Matthias Pickem, Motoharu Kitatani, Anna Kauch, Karsten Held},
  pdftitle = {Self-consistent ladder DGA approach},
  pdfsubject = {dgasc},
  pdfkeywords = {DGA},
  pdfdisplaydoctitle = true,
  colorlinks=true,urlcolor=blue,citecolor=blue,linkcolor=blue,breaklinks=true
}

\begin{document}
\title{Self-consistent ladder \bfdga\ approach}
\author{Josef Kaufmann}
\affiliation{Institute of Solid State Physics, TU Wien, 1040 Vienna, Austria}
\author{Christian Eckhardt}
\affiliation{Institute for Theoretical Solid State Physics, RWTH Aachen University, 52074 Aachen, Germany}
\affiliation{Institute of Solid State Physics, TU Wien, 1040 Vienna, Austria}
\author{Matthias Pickem}
\affiliation{Institute of Solid State Physics, TU Wien, 1040 Vienna, Austria}
\author{Motoharu Kitatani}
\affiliation{RIKEN Center for Emergent Matter Sciences (CEMS), Wako, Saitama, 351-0198, Japan}
\author{Anna Kauch}
\affiliation{Institute of Solid State Physics, TU Wien, 1040 Vienna, Austria}
\email{kauch@ifp.tuwien.ac.at}
\author{Karsten Held}
\affiliation{Institute of Solid State Physics, TU Wien, 1040 Vienna, Austria}
\date{\today}

\begin{abstract}
We present and implement a self-consistent \dga\ approach for multi-orbital models and {\em ab initio} materials calculations. It is applied to the one-band Hubbard model at various interaction strengths with and without doping,  to the two-band Hubbard model with two largely different bandwidths, and to SrVO$_3$.
 The self-energy feedback reduces critical temperatures compared to  dynamical mean-field theory, even to zero temperature in two-dimensions. Compared to a one-shot, non-self-consistent calculation the  non-local correlations are  significantly reduced when they are strong. In case non-local correlations are weak to moderate as for SrVO$_3$, one-shot calculations are sufficient.
\end{abstract}

\keywords{DGA}

\maketitle

\section{Introduction}
Strongly correlated materials are becoming more and more relevant for technological applications.   They are also utterly fascinating, not least because 
their theoretical study is intrinsically difficult. The actual calculation of correlated materials and their properties 
 usually requires a combination of \abinitio \ methods and simplified model approaches.
A very  successful \abinitio \ method for studying strongly correlated materials
is the combination of density functional theory \cite{Hohenberg1964,Jones1989} with the dynamical mean-field theory~\cite{PhysRevLett.62.324,Georges1992,Jarrell1992,RevModPhys.68.13,kotliar2004strongly}
(DFT + DMFT)~\cite{RevModPhys.68.13, Anisimov1997, Lichtenstein1998, kotliar2004strongly, held06, Kotliar2006, held2007electronic}, 
which is capable of describing local electronic correlations very accurately. 
In systems where nonlocal correlations play an important role, 
e.g., in two-dimensional or layered systems, DMFT cannot predict the correct low temperature behavior.
Cluster and diagrammatic extensions of DMFT~\cite{Gull2011,rohringer18rmp} have been developed to cure this problem. 

One such method is the \abinitiodga\ ~\cite{galler17, Galler18, galler19} which extends the concept of the dynamical vertex approximation (D$\Gamma$A) \cite{original_dga,Katanin2009} to realistic materials calculations. It inherits from DMFT the non-perturbative treatment of strong local correlations, but on top of this also includes non-local correlations. To this end, 
a two-particle ladder is built with the local DMFT irreducible vertex and the non-local Green's function as building blocks.
These ladder diagrams then yield a  non-local contribution to the self-energy.

Hitherto such \abinitiodga\ calculations have been restricted to so-called ``one-shot'' calculations without an update of the DMFT vertex and non-local Green's function. Obviously, such a one-shot calculation is only expected to be reasonable as long as the non-local corrections to DMFT remain small. It also does not suppress the DMFT critical temperatures nor modifies the DMFT critical exponents.
In the case of D$\Gamma$A calculations for one-band models, so-far 
a Moriyaesque $\lambda$-correction~\cite{Katanin2009, Rohringer2016} was devised as a cure.
It imposes a sum rule on the spin (or alternatively spin and charge) susceptibility, reduces the critical temperature and yields reasonable critical exponents\cite{Rohringer2011,Schaefer2017,Schaefer2019}.  Superconductivity in  cuprates~\cite{Kitatani2019} and nickelates~\cite{Kitatani2020} is described surprisingly accurately, even correctly predicted in the latter case.
The extension to the multi-orbital case however makes this Moriyaesque $\lambda$-correction difficult. One would need to introduce and determine various  $\lambda$ parameters for all orbital combinations and spin channels. This would result in a multi-dimensional optimization problem that is likely to have several local optima of comparable quality; details of how the
$\lambda$-correction is modeled (which is nonunique) might be decisive.

Another route has been taken in the closely related dual fermion approach \cite{Rubtsov08} with ladder diagrams \cite{Hafermann2009}. Here,  the Green's function is updated with the calculated non-local self-energy in a so-called ``inner self-consistency''. Hitherto applied to one-band model Hamiltonians such as the Hubbard \cite{Hirschmeier2015} and Falicov-Kimball model \cite{Antipov2014} yields very reasonable critical temperatures and exponents. Also a self-consistent update of the dual fermion vertex has been discussed \cite{Ribic2018, Loon2018, Tanaka18}.

In case of the \dga\ such an update of the Green's function has also been made, however only for the much more involved parquet \dga~\cite{Valli2015a, Li16, Li19, Kauch2019,Eckhardt20}\footnote{For the parquet dual fermion approach,  see \cite{Astleithner20, Astretsov19, Krien20}}.   Here, besides the self-consistent update of the Green's function and self-energy,
all three scattering (ladder) channels are mutually fed back into all other channels through the parquet equation  \cite{DeDominiciesMartin64a, DeDominiciesMartin64b, Bickers04, Vasiliev98}.
The drawback is the extreme numerical effort needed to solve the parquet equations,
which limits the method to one-band models so far~\cite{Valli2015a,Li16,Kauch2019, Kauch2020}.

In this paper we present a self-consistent ladder \dga \ (sc-\dga) for multi-orbital models. We update the  Green's function lines, 
as it is also done in parquet and dual fermion approaches but neither in the original \abinitiodga \ method nor in previous ladder \dga \ calculations.
This allows for a self-energy feedback into the ladder diagrams contained in the Bethe-Salpeter equation,
and leads to substantial damping of the fluctuations in the respective scattering channel.
Since this approach only requires a repeated evaluation of the  \abinitiodga \ equations, 
its application to multi-orbital models is straightforward.    Our results demonstrate that sc-\dga \ works well for single- and multi-orbital systems and also when doping away from integer filling.

The paper is organized as follows:
In \cref{sec:modelformalism} we introduce the Hubbard model (HM),
our notation, and the DMFT. 
Furthermore we give an overview over the different variants of
D$\Gamma$A that were hitherto used. 
In \cref{sec:scdga} we introduce our new way of doing D$\Gamma$A
self-consistently.
Then, in \cref{sec:results-one}, we present results for the
single-orbital Hubbard model on the square lattice with nearest-neighbor
hopping. This model has already been extensively studied
and our results can be compared to the literature.
Finally, in \cref{sec:results-multi}, we present results
for a two-orbital model system with Kanamori interaction,
and for SrVO$_3$ at room temperature.

\section{Model and formalism}
\label{sec:modelformalism}
\subsection{Multi-orbital Hubbard model}
The Hamiltonian of the multi-orbital Hubbard model reads
\begin{align}
  \label{eq:hubbard-model}
  H_\text{HM} = & \sum_{\vect{k}} \sum_{lm\sigma} 
	h_{lm}(\vect{k}) \cdag_{\vect{k}l\sigma} \cee_{\vect{k}m\sigma}\notag\\
	&+ \sum_i \sum_{\substack{ll'mm'\\ \sigma\sigma'}} U_{lm'ml'} \cdag_{im'\sigma} \cdag_{il\sigma'} \cee_{im\sigma'} \cee_{il'\sigma} \; .
\end{align}
Here, the first term is the
underlying tight-binding model, which can be obtained \emph{ab initio} by Wannierization
of a bandstructure from density functional theory. 
The operator $\cdag_{\vect{k} m \sigma}$ ($\cee_{\vect{k} m \sigma}$) creates (removes) an electron
with spin $\sigma$ in the Wannier orbital $m$ at momentum $\vect{k}$  (the Fourier transformed operators are labeled with unit cell index  $ i$ instead of $\mathbf{k}$).
The second term of \cref{eq:hubbard-model} contains
the interaction of the electrons. While the underlying \abinitiodga\ can in principle include non-local interactions, 
we here restrict ourselves to local ones.  That is, in each unit cell $i$, the matrix $U_{lm'ml'}$
parameterizes scattering events in which local orbitals $l,l',m,m'$
are involved.
In cases where the unit cell contains multiple atoms,
the matrix elements of $U_{lm'ml'}$ are non-zero only
when all indices correspond to interacting orbitals
of the same atom (i.e., are  local interactions). This restriction can be relaxed, in principle, to include also non-local interactions within the unit cell, either defining the whole unit cell as ``local'' or including the bare non-local interactions within the (then non-local)  vertex building block for ladder \dga.

The physics of the Hubbard model is usually studied in the framework
of the Green's function formalism. Our computational methods
additionally employ the Matsubara formalism, where the one-particle Green's function
for a system in thermal equilibrium at temperature $T=1/\beta$ is defined by 
\begin{equation}
  \label{eq:1pgf-hubbard}
	G_{lm}^k = -\int_0^\beta \hspace{-5pt} d\tau \,e^{i\nu\tau}\; \langle \wick \cee_{\vect{k}l}(\tau)
							  \cdag_{\vect{k}m}(0) \rangle.
\end{equation}
Here, the 4-index $k=(i\nu,\vect{k})$ combines Matsubara frequency $i\nu$ and crystal momentum
$\vect{k}$; $\tau$ is the imaginary time. Spin indices were omitted here, 
since we consider only paramagnetic systems with spin-diagonal Green's functions.
The interacting Green's function contains (infinitely) many connected
Feynman diagrams that are, via the Dyson equation (DE), captured by the self-energy:
\begin{equation}
  \label{eq:dyson}
  \Sigma_{lm}^k 
  = (i\nu + \mu)\delta_{lm} 
    - h_{lm}(\vect{k}) 
    - \big[G^k\big]^{-1}_{lm} \; .
\end{equation}

\subsection{Dynamical mean-field theory}
In most cases, it is completely infeasible to compute $G_{lm}^k$
or $\Sigma_{lm}^k$ directly through these infinitely many Feynman diagrams. Instead, one is bound to rely on approximations.
In the DMFT approximation the self-energy is assumed to be strictly local, or momentum-independent.
This becomes exact in infinite dimensions, while it still remains an excellent approximation
in three dimensions, and even for many two-dimensional systems. 
\begin{figure}
  \centering
  \includegraphics[width=0.5\textwidth]{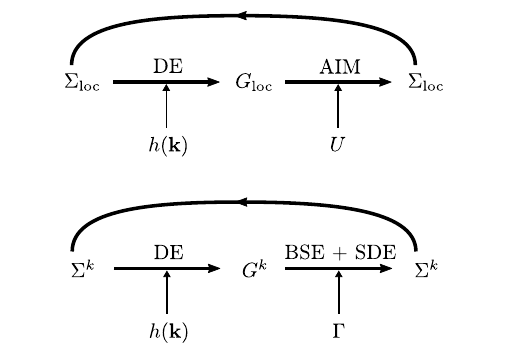}
  \caption{\label{fig:flowchart}Schematic explanation of DMFT and \dga\ loops.}
\end{figure}
As we illustrate in \cref{fig:flowchart} in a very abstract way,
DMFT consists of two steps: First, one uses the $\mathbf{k}$-integrated Dyson equation  (\ref{eq:dyson})
to obtain the local Green's function from the local ($\vect{k}$-independent) DMFT self-energy:
\begin{equation}
  \label{eq:dyson-loc}
  G_{lm}^\nu
  = \frac{1}{V_\text{BZ}} \int_\text{BZ} \dif^d{k}
  \big[
  (i\nu + \mu)\delta_{lm} 
    - h_{lm}(\vect{k}) 
    - \Sigma_{lm}^{\nu} \big]^{-1}
\end{equation}
Here, the integral over the crystal momentum
$\vect{k}$ is taken over the first Brillouin zone (BZ) with volume ${V_\text{BZ}}=(2\pi)^d/V$  ($V$: unit cell volume; $d$: dimension).
The chemical potential $\mu$ is chosen such that the system
contains the desired number of electrons.
In the second step one obtains a new local self-energy,
which is in principle the sum of all self-energy diagrams
built from the above propagator and the local interaction.
These two steps can be iterated until convergence.

In practice the second step is usually solved by introducing an auxiliary
Anderson impurity model (AIM), since a direct summation of all diagrams is infeasible.
For the AIM, on the other hand, it is possible to calculate correlation
functions like the one-particle Green's function $g_l^\nu$ on the impurity numerically exactly.

\subsection{Local correlations on the two-particle level}
Despite the success of DMFT, additional efforts are necessary in order to access
also the momentum dependence of the self-energy.
There are several diagrammatic extensions of DMFT that result in the momentum dependent self-energy (for a review see Ref.~\onlinecite{rohringer18rmp}). These diagrammatic routes to non-local correlations all rely on two-particle vertices from DMFT. 
Here locality is assumed on the two-particle level, 
instead of the one-particle level.
Local correlations on the two-particle level \cite{rohringer12} are contained
in the two-particle Green's function of the (DMFT) impurity model,
\begin{align}
  \label{eq:2pGF}
  G_{abcd}^{\nu_1 \nu_2 \nu_3 \nu_4} = \frac{1}{\beta^2}
  \hspace{-3pt}\int_0^\beta \hspace{-6pt}&\dif\tau_1\dif\tau_2\dif\tau_3\dif\tau_4 
  e^{i(\nu_1\tau_1 \!-\! \nu_2\tau_2 \!+\! \nu_3\tau_3 \!-\! \nu_4\tau_4)}\notag\\
  &\langle \wick \cee_a(\tau_1) \cdag_b(\tau_2) \cee_c(\tau_3) \cdag_d(\tau_4) \rangle,
\end{align}
for which we use spin-orbital compound indices $a$, $b$, $c$, $d$.
In this paper we compute such two-particle Green's functions
by continuous-time quantum Monte Carlo (CT-QMC)
with worm sampling \cite{Gunacker2015}, which is implemented in 
{\sc w2dynamics}~\cite{wallerberger2019w2dynamics}.

The two-particle Green's function is connected 
to the full reducible vertex $F_{abcd}$ by
\begin{equation}
  \label{eq:rel-2pgf-vertex}
  G_{abcd}^{\nu_1 \nu_2 \nu_3 \nu_4} = 
  g_a^{\nu_1} g_c^{\nu_3} \big( \delta_{12} - \delta_{14} \big)
  -\frac{1}{\beta} g_a^{\nu_1} g_b^{\nu_2} g_c^{\nu_3} g_d^{\nu_4}
  F_{abcd}^{\nu_1 \nu_2 \nu_3 \nu_4},
\end{equation}
where $\delta_{12} \equiv \delta_{ab} \delta_{cd} \delta_{\nu_1\nu_2}$
and $\delta_{14} \equiv \delta_{ad} \delta_{bc} \delta_{\nu_1\nu_4}$.
Closely related is the generalized susceptibility
\begin{align}
  \label{eq:gen-susc}
  \chi_{abcd}^{\nu_1 \nu_2 \nu_3 \nu_4} &= \beta\Big(
    G_{abcd}^{\nu_1 \nu_2 \nu_3 \nu_4} - g_a^{\nu_1} g_c^{\nu_3} \delta_{12} \Big)\\
    &\equiv \chi_{0,abcd}^{\nu_1 \nu_2 \nu_3 \nu_4} 
    + \chi_{\text{conn},abcd}^{\nu_1 \nu_2 \nu_3 \nu_4},
\end{align}
with
\begin{equation}
  \label{eq:chi0}
  \chi_{0,abcd}^{\nu_1 \nu_2 \nu_3 \nu_4} = -\beta g_a^{\nu_1} g_c^{\nu_3} \delta_{14}.
\end{equation}
Since energy conservation constrains $\nu_1 + \nu_3 = \nu_2 + \nu_4$,
it is sometimes of advantage~\footnote{And often dangerous!}
to make a transition from four fermionic frequencies
to a notation with two fermionic and one bosonic Matsubara frequency.


If we choose the bosonic frequency as $\omega_\text{ph} = \nu_1 - \nu_2$,
the Bethe-Salpeter equation (BSE) in the particle-hole channel
can be solved separately at each bosonic frequency.
In the particle-particle channel, we have to choose 
$\omega_\text{pp} = \nu_1 + \nu_3$ instead.
Furthermore, the Bethe-Salpeter equations can be diagonalized in spin space
by the following linear combinations:
\begin{align}
  F_{\mathrm{d}, nlhm} 
    &= F_{n\uparrow l\uparrow h\uparrow m\uparrow} 
      + F_{n\uparrow l\uparrow h\downarrow m\downarrow} \; ,\\
  F_{\mathrm{m}, nlhm} 
    &= F_{n\uparrow l\uparrow h\uparrow m\uparrow} 
      - F_{n\uparrow l\uparrow h\downarrow m\downarrow} \; ,\\\
  F_{\mathrm{s}, nlhm} 
    &= F_{n\uparrow l\uparrow h\uparrow m\uparrow} 
      + F_{n\uparrow l\uparrow h\downarrow m\downarrow} \; ,\\\
  F_{\mathrm{t}, nlhm} 
    &= F_{n\uparrow l\uparrow h\uparrow m\uparrow} 
      - F_{n\uparrow l\uparrow h\downarrow m\downarrow} \; .
\end{align}
The Bethe-Salpeter equations for the impurity in the particle-hole (ph) channel are thus
\begin{equation}
  F_{r,lmm'l'}^{\nu\nu'\omega} = \Gamma_{r,lmm'l'}^{\nu\nu'\omega}
  + \sum_{\substack{nn'hh'\\\nu''}} \Gamma_{r,lmhn}^{\nu\nu''\omega} 
    \chi_{0, nhh'n'}^{\nu''\nu''\omega} 
    F_{r,n'h'm'l'}^{\nu''\nu'\omega}
\end{equation}
where $r=d,m$ denotes the afore-defined channel and $\omega\equiv\omega_{ph}$ is
the bosonic frequency. 
For better readability we will adopt the shorthand notation
\begin{equation}
  \label{eq:bseq-impurity}
  F_r^\omega = \Gamma_r^\omega + \Gamma_r^\omega \chi_0^\omega F_r^\omega,
\end{equation}
where all quantities are matrices in an orbital-frequency compound index.

\subsection{Dynamical vertex approximation}
The \dga\ is a diagrammatic extension of DMFT that assumes locality of the irreducible vertex,
which is taken as input from an auxiliary impurity problem 
(usually from a converged DMFT solution to the original problem). 

Since its original formulation in Ref.~\onlinecite{Toschi2007}, 
the \dga\ was developed in three main directions (often called different \dga\ {\it flavors}): 
\begin{itemize}
\item[(i)] the original parquet formulation (p-\dga), 
  where the locality is assumed on the level of the fully irreducible 
    vertex $\Lambda$ -- this flavor treats the smallest set of diagrams as local,
    and correspondingly it is computationally most demanding~\cite{Li19,Eckhardt20}; 

\item[(ii)] ladder-\dga \ (in combination with DFT input also called \abinitiodga \ \cite{galler17, Galler18, galler19}), 
  where it is the irreducible vertex in the particle-hole channel ($\Gamma_{ph}$) that is assumed local; 

\item[(iii)] $\lambda$-corrected \dga \ (usually also called ladder-\dga), 
  where as in (ii) the irreducible vertex $\Gamma_{ph}$ is taken as local.
    However, after the solution of the Bethe-Salpeter equations
    (a non-local version of  \cref{eq:bseq-impurity}),
    a sum rule is imposed on the susceptibility by introducing the so-called Moriyaesque $\lambda$-correction~\cite{Katanin2009,  Rohringer2016,Schaefer2016}
    to the susceptibility and self-energy. 
\end{itemize}
\noindent Below we first briefly review these three existing flavors, as this allows  placing the new flavor (sc-\dga; introduced in the next Section)
into its proper methodological context. 

\subsubsection{Parquet-\itdga}
The parquet-scheme is a method to self-consistently calculate 1-particle and 2-particle quantities \cite{DeDominiciesMartin64a, DeDominiciesMartin64b, Bickers04, Vasiliev98} (it is closely related to the multiloop generalization \cite{Kugler18, Hille20} of the functional renormalization group (fRG) method \cite{Metzner12}).
Given a one-particle Green's function $G$ and the fully 2-particle irreducible~\footnote{That is, cutting any two Green's function lines does not separate the Feynman diagram into two pieces.} 2-particle vertex $\Lambda$, one can iterate the parquet equation
\begin{equation}
	F_{r}= \Lambda_{r} + \sum_{r'} c_{r'} \underbrace{\Gamma_{r'} \chi_0 F_{r'}}_{\Phi_{r'}}
	\label{eq:latticeParquet}
\end{equation}
and the lattice BSE
\begin{equation}
	F_{r}^q = \Gamma_{r}^{q} + \Gamma_r^{q} \chi_0^{q} F_r^{q}
	\label{eq:latticeBSE}
\end{equation}
to obtain the (in general non-local) vertices $F_{r}$ and $\Gamma_{r}$.
Here the index $r = d,m,s,t$ is as defined earlier the channel index, $c_r$ denotes a real prefactor, and~\cref{eq:latticeBSE} is diagonal in the bosonic variable (4-index $q$). In our short notation $F^{q}_{r}$ and $\Gamma^{q}_{r}$ are matrices 
 in two fermionic multi-indices as before. The parquet equation~\cref{eq:latticeParquet} is not diagonal in the bosonic 4-index and its evaluation requires evaluation of the per definition reducible vertices $\Phi_{r}$ at different frequency and momentum combinations (for explicit formulation see e.g.~Ref.~\onlinecite{Li19}).

The Green's function entering the above Eqs.~\eqref{eq:latticeParquet}-\eqref{eq:latticeBSE} via $\chi_0$ can also be updated, since the full vertex $F$ is related to self-energy through the Schwinger-Dyson equation (SDE, see e.g.~Ref.~\onlinecite{Li19}). 

The SDE together with the Dyson equation and Eqs.~\eqref{eq:latticeParquet}-\eqref{eq:latticeBSE} 
constitute a closed set with only one input quantity: $\Lambda$. For an exact $\Lambda$, the parquet scheme produces the exact one- and two-particle quantities. In practice, for example  $\Lambda = U$ is taken, which is the lowest order in perturbation expansion widely known
as the parquet approximation~\cite{DeDominiciesMartin64b, Bickers04}. In the parquet \dga\ method  $\Lambda$ is assumed local and taken from a converged DMFT calculation~\footnote{It is not necessary to take  $\Lambda$ from DMFT. One could improve the auxiliary Anderson impurity model, so that it produced the same local Green's function as the one resulting from p-\dga. It would add another level of self-consistency. This impurity update turned out to be unnecessary for the parameters presented in this paper.}.

{\it Truncated unity approximation.}
The parquet scheme is numerically extremely costly \cite{Li19}.
We thus employ an additional approximation. Specifically,  
we transform the fermionic momentum dependence of the 2-particle reducible vertices $\Phi_r$ into a real space basis, leaving only the bosonic momentum ${\bf q}$:
\begin{equation}
	\tilde{\Phi}^{\ell \ell' q}_{r}%
	= \frac{1}{N} \sum_{k, k'} ({f^{k \ell}})^* \Phi^{k k' q}_{r} f^{k' \ell'}
	\label{eq:formfactorTransformation}
\end{equation}
where $f^{k \ell}$ are basis functions (typically known as form factors) of a suitable transformation-matrix which we choose to obey certain symmetries.
Exploiting the relative locality \cite{Husemann09, Eckhardt20} of the reducible vertices $\Phi$ in their two fermionic momenta we limit the number
of basis functions $f^{k \ell}$ used for the transformation (hence the name {\it truncated unity}).
This amounts to setting the more nonlocal parts (in the fermionic arguments) of the 2-particle reducible vertices to zero.
\begin{equation}
	\tilde{\Phi}^{\ell \ell' q}_r= 0 \hspace{2mm} \text{for} \hspace{2mm} \ell, \ell' > l_{\text{max}}.
	\label{eq:approxmationConditionFormfactors}
\end{equation}
The calculations to transform the entire parquet-scheme including convergence studies in the number of basis functions
can be found in Refs.~\onlinecite{Eckhardt18, Eckhardt20}. The truncated unity implementation (TUPS)~\cite{Eckhardt20} with 1 or 9 form factors was used to generate the comparison data in Sec.~\ref{sec:results-one}.


\subsubsection{Ladder-\itdga}
Even with the truncated unity approximation the parquet-\dga\ is numerically very costly. It also suffers
from the presence of divergencies~\cite{Schaefer2013, Schaefer2016PRB, Gunnarsson2017, Vucicevic2018, Chalupa2018} in the fully irreducible vertex $\Lambda$ that is directly taken as input. Therefore  
it is often preferable to use ladder-\dga, where the locality
level is raised to the irreducible vertex in the particle-hole channel $\Gamma_{d/m, \text{imp}}$. The choice of channel is here determined by the dominant type of non-local fluctuations. By choosing the particle-hole channel we take non-local magnetic and density fluctuations into account, while treating particle-particle fluctuations only at the local level.~\footnote{In some cases it may be preferable to take the non-local particle-particle fluctuations as dominant and approximate the particle-hole channel to the local level. In the scope of this paper we treat however only problems with dominant fluctuations in the particle-hole channel.}  Note that the transversal particle-hole fluctuations will later be included on the same level by using the crossing symmetry, which relates it to the particle-hole channel.

Then, \dga\ becomes significantly simpler and essentially consists of two
steps: First one has to compute the Bethe-Salpeter equations
\begin{equation}
  \label{eq:bseq-ladder-dga-1}
  F_r^q = \big[1 - \Gamma_{r,\text{imp}}^\omega \chi_{0}^q \big]^{-1} \Gamma_{r,\text{imp}}^\omega 
\end{equation}
in the particle-hole channels $r=\mathrm{d},\mathrm{m}$.
Since the irreducible impurity vertices $\Gamma_{r,\text{imp}}$
can also exhibit divergences, it is better
to reformulate the above equation. This is done by expressing $\Gamma$
by \cref{eq:bseq-impurity} and rearranging the terms, as shown
in Ref.~\onlinecite{galler17}. Then one arrives at
\begin{equation}
  \label{eq:bseq-ladder-dga}
  F_r^q = F_r^\omega \big[1 - \chi_{0}^{\text{nl},q} F_r^\omega \big]^{-1}
\end{equation}
containing only the full reducible vertex $F$, and the non-local part of the bubble
$\chi_{0}^{\text{nl},q} = \chi_0^q - \chi_0^\omega$.

The momentum-dependent reducible vertices $F_r^q$ from the 
longitudinal and transversal particle-hole channels are then combined.
We do not need to calculate the latter explicitly, 
because it can be obtained from the former through the crossing-symmetry \cite{rohringer18rmp}.
The combined vertex $\mathcal{F}$ is then
\begin{align}
  \mathcal{F}_{\mathrm{d},nlhm}^{kk'q}
  =
  &F_{d,nlhm}^{\nu\nu'\omega}
  + F_{d,nlhm}^{\text{nl},\nu\nu'q}\notag\\
  &- \frac{1}{2} F_{d,hlnm}^{\text{nl},(\nu'-\omega)\nu'(k'-k)}
  - \frac{3}{2} F_{m,hlnm}^{\text{nl},(\nu'-\omega)\nu'(k'-k)}
\label{eq:crossing_symmetry}
\end{align}
(see also Eq.\ (54) in Ref.\ \onlinecite{galler17}). Vertices labeled ``nl''
are non-local, i.~e.~$F_{r,nlhm}^{\text{nl},\nu\nu'q} = F_{r,nlhm}^{\nu\nu'q} - F_{r,nlhm}^{\nu\nu'\omega}$.
Inserting this into the Schwinger-Dyson equation of motion~\cite{galler17}
\begin{equation}
  \label{eq:schwidy}
  \Sigma_{mm'}^{\text{con},k} = -\frac{1}{\beta}\hspace{-5pt}\sum_{nlhn'l'h'} \sum_{k'q} 
  U_{mlhn}\chi_{0, nll'n'}^{k'k'q}\mathcal{F}_{\mathrm{d},n'l'h'm'}^{kk'q}G_{hh'}^{k-q}
\end{equation}
yields the connected part of the momentum-dependent self-energy.
In practice this equation is evaluated separately for the summands of $\mathcal{F}$ in Eq.~(\ref{eq:crossing_symmetry})~\cite{galler19},
such that one can identify the non-local corrections to the DMFT self-energy.

Eqs.\ \eqref{eq:bseq-ladder-dga}-\eqref{eq:schwidy} can be evaluated efficiently 
even for multi-orbital models with $h(\mathbf{k})$ from DFT as input. 
This is known as the \abinitiodga\ ~\cite{galler17, Galler18, galler19}. 
Hitherto they are evaluated only once, 
and this flavor is therefore referred to as one-shot \dga \ ($1$-\dga) in the following.

\subsubsection{$\lambda$-corrected \itdga}
The self-energy obtained in the one-shot ladder-\dga \ calculation does
not always exhibit the correct asymptotic behavior,
especially if the susceptibility is large. In addition, the susceptibilities related to \cref{eq:bseq-ladder-dga}
diverge at the DMFT N\'eel temperature, violating
the Mermin-Wagner theorem \cite{Mermin1966} for 2-dimensional models.  
This problem was partially solved by so-called $\lambda$-corrections~\cite{Katanin2009, Rohringer2016},
where one enforces the sum rule for the spin (or spin and charge) susceptibility(-ies)
by adapting a parameter $\lambda$ (hence the name).

While very successful for one-band models ~\cite{Schaefer2016,Schaefer2017,Schaefer2019,Kitatani2019,Kitatani2020,schaefer20}, 
this solution is not straightforwardly extensible to multi-orbital systems. The reasons are twofold. 
Firstly, $\lambda$ would be a matrix with as many independent entries as there are different spin-orbital combinations,
resulting in a multi-dimensional optimization problem. 
Secondly, the solution to this problem is quite likely nonunique
and there are at the moment no criteria how the physical matrix $\lambda$ should be chosen. 
While we do not exclude that a reasonable scheme can be devised 
for the multi-orbital case in the future (see e.g.~\cite{TPSC_Arita, TPSC_Tremblay} for application of sum rules in 
the multi-orbital two-particle self-consistent approach~\cite{Tremblay2011}),
we focus here on an alternative scheme, 
that does not rely on enforcing sum rules.

\section{Self-consistent ladder \bfdga}
\label{sec:scdga}

While the $\lambda$-correction is 
impractical or perhaps not even possible for multi-orbital systems, 
a one-shot ladder-\dga \ calculation as hitherto employed for realistic materials calculations also has severe limits.
Where the non-local corrections become strong, its application is not justified. 
When the DMFT susceptibility diverges at a phase transition, 
the non-local corrections of a one-shot \dga\ calculation are not meaningful any more.

There are two main physical reasons why this is wrong:
Firstly, the ladder diagrams of say the particle-hole channel lack insertions from the particle-particle channel, 
which dampen the particle-hole fluctuations. 
These diagrams are taken into account only on the level of the impurity.
In order to correctly incorporate the non-local contributions to such insertions,
we need to evaluate the full parquet scheme
that is at the moment numerically too costly for multi-orbital calculations. 

Secondly and arguably even more important, the self-energy that enters the propagators in the BSE 
is still the local DMFT self-energy in a one-shot \dga. 
This DMFT self-energy fulfills the local SDE with local $F^\omega$, 
where the non-local contributions do not enter. 
By using the updated non-local self-energy in the BSE,
we can introduce feedback from two-particle non-local correlations 
to the one-particle quantities.  For example, spin fluctuations lead to a reduced life-time which, when included in the ladder Green's function or self-energy, reduces the spin fluctuations in turn. This mechanism hence suppresses the magnetic transition temperature below the DMFT mean-field value.

\subsection{sc-\bfdga}

The approach we propose here consists in
finding a momentum-dependent self-energy
for a lattice, defined by the tight-binding Hamiltonian $h(\vect{k})$, 
that is consistent with the local irreducible vertex $\Gamma_{ph,\text{imp}}$. 
This can be achieved by using an iterative scheme illustrated in  
the lower panel of \cref{fig:flowchart} in order to underline its
formal similarity to DMFT: The first step is again the construction of propagators by the
DE [\cref{eq:dyson}], with a chemical potential that constrains the electron number.
But in contrast to DMFT the self-energy is now momentum-dependent.
In the second step we sum up all self-energy diagrams
that are generated from the local vertex $\Gamma$.
More explicitly, this step consists of the subsequent evaluation
of the BSE [\cref{eq:bseq-ladder-dga}] and SDE [\cref{eq:schwidy}].
Just as in DMFT, also here the second step is numerically much more expensive
than the first (DE) step.

The self-energy resulting from the first iteration of ladder-\dga\ 
is taken to be the input (or ``trial'') for the second iteration.
Starting from the third iteration, linear combinations of
trial and result self-energies from several previous iterations are
used as new trials. The linear combination is constructed
by the Anderson acceleration algorithm~\cite{Anderson1965,Walker2011}; see also Appendix~\ref{sec:implementation}.
If the result is equal to the trial, the iteration is stopped.
The workflow of such a calculation is illustrated in \cref{fig:algorithm}.

To our knowledge there is no proof of uniqueness or existence of such
a fixed point. However, we find the procedure to be convergent over a
large range of parameters (cf.~\cref{fig:phase-diagram} which is discussed in Sec.~\ref{sec:results-one}).

In case of convergence, the asymptotic behavior of the self-energy
is largely repaired with respect to one-shot \dga \ calculations.
Furthermore, the magnetic susceptibility in two-dimensional models stays finite at all 
temperatures in agreement with the Mermin-Wagner theorem.

\subsection{Implementation and computational effort}

\begin{figure}
  \includegraphics[width=0.5\textwidth]{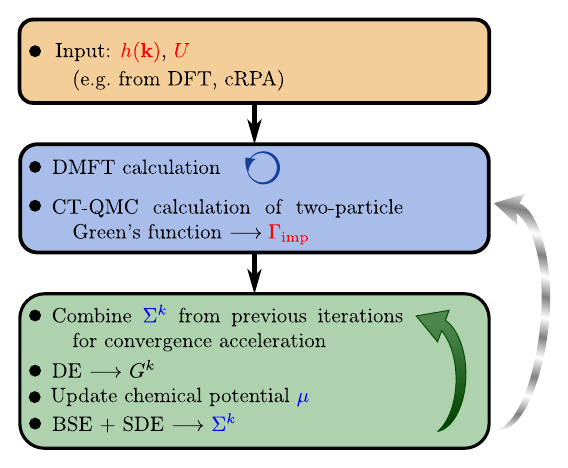}
  \caption{\label{fig:algorithm} Step-by-step illustration of a self-consistent ladder \dga\ calculation.
  The first box (orange color) shows preliminary calculations to set up the model. 
  The second box (blue color) concerns the DMFT calculations, and the third one (green color)
  shows the \scdga-cycle. Quantities written in red are kept constant throughout the whole calculation;
  those in blue are self-consistently determined in \scdga.}
\end{figure}

The sc-\dga \ is applicable to multi-orbital calculations using the AbinitioD$\Gamma$A  code~\cite{galler19},
with the slight modification of allowing for momentum-dependent self-energies in the input.
A step-by-step description of the workflow is given in \cref{fig:algorithm},
whereas in \cref{sec:implementation} we provide more technical details of how this is done
in operation with the  AbinitioD$\Gamma$A.

The first step (orange box in \cref{fig:algorithm}) is the creation of a model.
It can be based on \abinitio\ calculations and consists of a tight-binding Hamiltonian
as well as a parametrization of the interaction in form of a $U$-matrix. 

The second step is to determine a local (impurity) vertex  $\Gamma_\text{imp}$.
Here, this is obtained  from the local impurity problem at  DMFT self-consistency
as indicated in the blue box in \cref{fig:algorithm}; then
usually the DMFT self-energy is also taken as a starting point for the following \dga\ calculations.
It is however not strictly required to start from a converged DMFT
calculation. One might as well start from $GW$ or calculate  $\Gamma_\text{imp}$
from the \dga\ Green's function in an additional self-consistency step, as indicated by the dashed gray arrow
in  \cref{fig:algorithm} but not done in the this paper.

Finally, the actual \scdga\ cycle is illustrated in the green box of \cref{fig:algorithm}. 
It essentially amounts to the execution of the  AbinitioD$\Gamma$A  code~\cite{galler19},
but in the repeated evaluation of Eqs.~\eqref{eq:bseq-ladder-dga}-\eqref{eq:schwidy},
we have to generate updated input quantities after every iteration,
until convergence is reached. Here, also the chemical potential $\mu$ is adjusted so that the total number of electrons is kept fixed.

In the \scdga\ implementation the local irreducible vertex is never used explicitly,
and the equations are evaluated in terms of the local full vertex $F^\omega$ 
(\cref{eq:bseq-ladder-dga} instead of \cref{eq:bseq-ladder-dga-1}). 
As already mentioned, this avoids  the computational difficulties coming from using a very large irreducible vertex 
near or on a divergence line. Indeed,
 the sc-\dga\ scheme can be converged also quite close to the divergence lines (cf.~\cref{fig:phase-diagram}).
Let us however note that the local part of self-energy in the converged sc-\dga\ 
calculation is in general not related to $F^\omega$ via the local SDE 
(as it was the case in a one-shot ladder-\dga). 
The sc-\dga \ corrections to the self-energy modify thus also 
its local part that is not any more equal to the DMFT solution.   
One can envisage~\cite{galler17,Ayral2016,rohringer18rmp} 
an update of the local multi-orbital vertex $\Gamma$ (dashed gray arrow in  \cref{fig:algorithm}; not implemented here) so that the local Green's function of the impurity is equal to the local sc-\dga \ Green's function. Such an update is at the moment numerically prohibitively expensive and hence beyond our scope.

At this point, it is appropriate to comment on the computational effort
of the present self-consistent ladder \dga.
The cost of a DMFT calculation and the two-particle Green's function
depends mainly on the desired accuracy, if one is using a Monte Carlo
method as an impurity solver. The scaling of the CT-QMC with temperature and number of 
orbitals has been discussed in the literature \cite{Gull2011}.
The measurement of the two-particle Green's function in worm sampling of w2dynamics
scales as $\sim(\#\omega)^3 M_\text{comp}$, where $M_\text{comp}$ is the number of
non-vanishing spin-orbital components. In the case of Kanamori interaction
this goes as $M_\text{comp}\sim M^2$, where $M$ is the number of impurity orbitals 
\footnote{Actually, for Kanamori interaction we find $M_\text{comp}(M) = 6 x_N$
where $x_N = 3M^2 - 2M$ are \emph{octagonal numbers}.}.
For a general dense $U$-matrix, the number of components is $M_\text{comp} = (2M)^4$.
The number of frequencies $\#\omega$ has to be scaled linearly with $\beta$
as lower temperatures are approached. Note that at this point we do not
distinguish between the number of fermionic and bosonic frequencies, since
at least their scaling with temperature is the same.

Having calculated the two-particle Green's function, the remaining
time of the computation is direct proportional to the number of iterations ($N_\text{iter}$)
needed for convergence. $N_\text{iter}$ ranges from a few ($\sim$10) at high-temperatures
to many ($\sim$200) iterations at low temperatures.
However, in problems with weak spin fluctuations, $N_\text{iter}$ is hardly dependent on temperature.
In our experience, convergence is accelerated if the DMFT self-energy $\Sigma$ used as a starting point
has little noise. 
Therefore we use symmetric improved estimators \cite{Kaufmann2019} to compute it in CT-QMC.
Noise in the vertex, on the other hand, does not have a large influence on the self-energy
in \dga, as shown recently \cite{Kappl2020}.

The computational effort of one \dga\ iteration has been discussed in Ref.\ \onlinecite{galler19}.
Let us give a brief overview for the sake of completeness.
In this part, most time has to be spent with the BSE, where it is necessary to
($\#\omega\#\mathbf{q}$) times invert a matrix of dimension ($M^2\#\omega$).
Overall this gives a scaling of $\sim\#\mathbf{q}(\#\omega)^{3.5}M^5$~\cite{galler19}.

In order to give a rough feeling or rule of thumb for the computational cost,
we remark that at high temperatures the DMFT and CT-QMC calculations take considerably more
time than the \dga\ self-consistency cycle.
For the most complicated cases, where many iterations are needed for convergence,
one may expect to spend about twice as much time for ladder \dga\ than
for the CT-QMC. We illustrate this by providing the actual CPU hours that
were spent on some of the calculations in \cref{tab:cpu-hours}.
\begin{table}
  \centering
  \begin{tabular}{c | r | r}
    Case & $[\textrm{Time}]_{\Gamma_{\rm imp}}$ & $\textrm{[Time}]_{\text{sc-D}\Gamma\text{A}}$\\
    \hline
    Sq.\ latt., $U=2$, $1/T=4$ & 2400$\,$h & 65$\,$h\\
    Sq.\ latt., $U=2$, $1/T=20$ & 43000$\,$h & 11000$\,$h\\
    2-band, $1/T=10$ & 13000$\,$h & 3000$\,$h \\
    2-band, $1/T=20$ & 40000$\,$h & 70000$\,$h
  \end{tabular}
  \caption{\label{tab:cpu-hours} Computational effort [time measured in core hours (h) on an Intel Skylake Platinum 8174 processors with 3.1~GHz] for the calculation of the local two-particle vertex $\Gamma_{\rm imp}$ (blue box in \cref{fig:algorithm})
      and the \scdga\  (green box in \cref{fig:algorithm}) for a few cases treated in this paper. ``Sq.\ latt.'' (square lattice) refers to \cref{sec:results-one},
  ``2-band'' refers to the two-band model treated in \cref{sec:results-multi}.}  
\end{table}

\subsection{Relation to p-\bfdga}
The self-consistency imposed on the self-energy that is obtained 
by iterative application of BSE \eqref{eq:bseq-ladder-dga}, 
crossing symmetry \eqref{eq:crossing_symmetry} and SDE \eqref{eq:schwidy} 
is reminiscent of the parquet scheme. 
The main difference is the lack of the full parquet equation~\eqref{eq:latticeParquet}, 
which would include also non-local particle-particle insertions in the full vertex $F$. 
In the full p-\dga \ the level of local approximation is also different, 
since  $\Lambda$ contains fewer diagrams than $\Gamma$. 
In the truncated unity approximation however, $\Gamma$ is also effectively 
local if we do calculations with only one form factor (1FF p-\dga). 
It can be explicitly seen e.g.\ in Eq.\ (21) in Ref.~\cite{Eckhardt20}. 
The difference between the irreducible vertices $\Gamma$
in the two approaches is that in sc-\dga \ it is taken from DMFT 
and never updated during the self-consistency cycle, 
whereas in 1FF p-\dga \ it is updated through the parquet equation in every iteration.
This update allows for mixing of scattering channels in 1FF p-\dga, 
notwithstanding the fact that the non-local contributions 
from other channels into $\Gamma$ are averaged over momenta.

\section{Square lattice Hubbard model}
\label{sec:results-one}
We begin the application of the sc-\dga\ method by considering a 
 relatively simple system,
which already has been studied well in some parameter regimes: the one-orbital
Hubbard model on a square lattice with nearest-neighbor hopping. 
The dispersion $h(\mathbf{k})$ in \cref{eq:hubbard-model} is then simply
\begin{equation}
  \label{eq:square-nn-dispersion}
  h(\mathbf{k}) = -2\big(\mathrm{cos}(k_x) + \mathrm{cos}(k_y) \big),
\end{equation}
where the nearest-neighbor hopping amplitude is set to $t\equiv 1$ to define our unit of energy
for this section (with $\hbar\equiv 1$ setting the frequency unit). Furthermore, the lattice constant $a\equiv 1$ sets the unit of length and $k_B\equiv 1$ the unit of temperature; and the orbital indices $l$, $m$, $l'$, $m'$ are restricted to a single orbital at each site.

\begin{figure}
  \centering
  \includegraphics[width=0.48\textwidth]{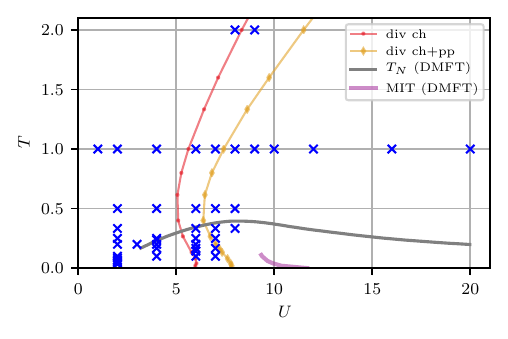}
  \caption{\label{fig:phase-diagram} Phase diagram of the square-lattice
    Hubbard model at half-filling. Blue crosses denotes points at which the sc-\dga \ could be converged. The DMFT-N\'eel temperature is
    shown in gray~(from \cite{Kunes11}). The magenta line indicates the DMFT metal-insulator transition. We also show the first two vertex divergence lines~(from \cite{Schaefer2013}).}
\end{figure}

\begin{figure}
  \includegraphics[width=0.5\textwidth]{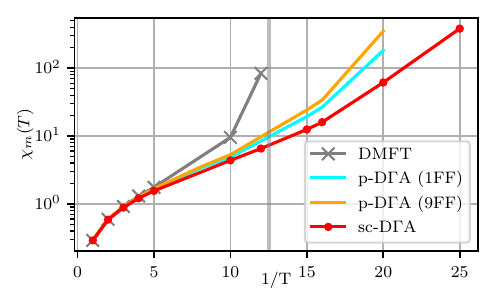}
  \caption{\label{fig:square-u2-chi} Static magnetic susceptibility 
    of the square-lattice Hubbard model with $U=2$ and $n=1$ at momentum 
    $\mathbf{q}=(\pi,\pi)$ as a function of inverse temperature. 
     Different colors and symbols denote different methods.
     The gray vertical line marks the DMFT N\'eel temperature.}
\end{figure}

\begin{figure*}
  \centering
  \includegraphics[width=\textwidth]{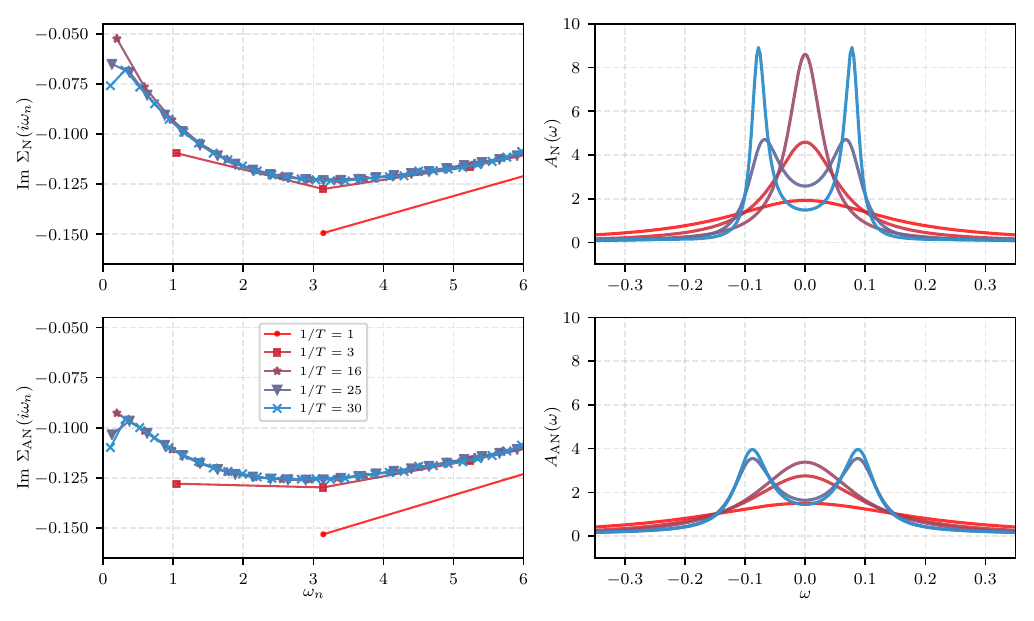}
  \caption{\label{fig:square-u2-se-scdga} The imaginary part of sc-\dga \ self-energy (left) and the corresponding spectral function (right) at $U=2$ and half-filling for two momenta on the Fermi surface: nodal point, $\mathbf{k}_{\mathrm{N}}=(\pi/2,\pi/2)$ and antinodal point, $\mathbf{k}_{\mathrm{AN}}=(\pi,0)$. Different colors and symbols denote different temperatures. The spectral functions were obtained by analytic continuation with the maximum entropy method~\cite{Geffroy2019,kaufmannGithub}. }
\end{figure*}

\begin{figure*}
  \includegraphics[width=\textwidth]{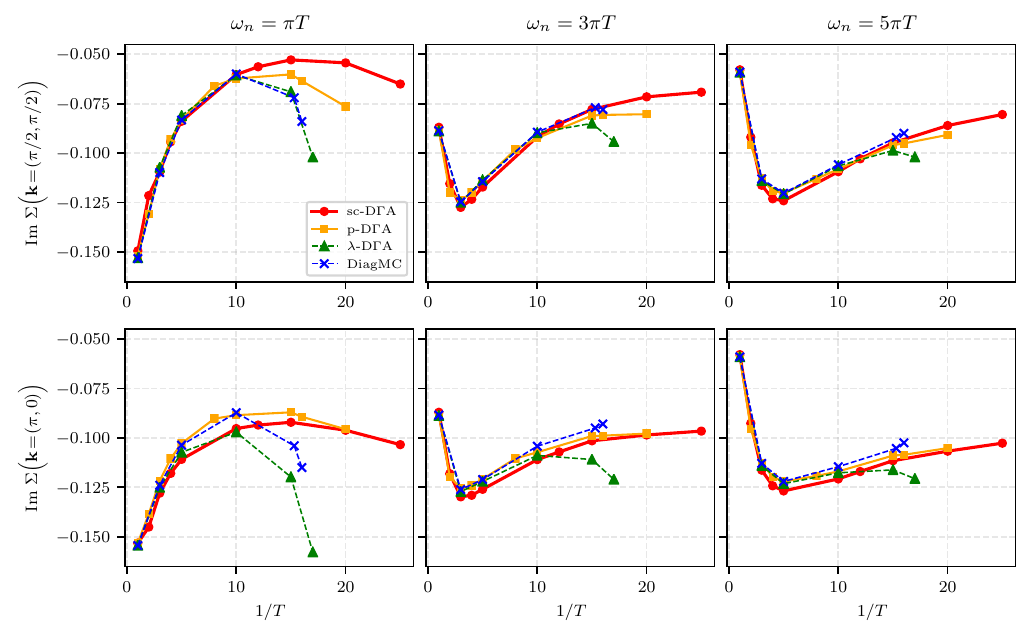}
  \caption{\label{fig:square-u2-se} 
  Inverse temperature dependence of the imaginary part of 
  self-energy at $U=2$ and half-filling for the first three Matsubara frequencies $\omega_n=\{\pi T,3\pi T, 5 \pi T\}$ for two momenta on the Fermi surface: $\mathbf{k}_{\mathrm{N}}=(\pi/2,\pi/2)$ and $\mathbf{k}_{\mathrm{AN}}=(\pi,0)$. Different colors and symbols denote different methods. The $\lambda$-\dga \ and DiagMC data in this figure were kindly provided by the authors of~\cite{schaefer20}.}
\end{figure*}

In~\cref{fig:phase-diagram} we show the DMFT phase diagram of the Hubbard model on 
a square lattice at half-filling ($n=1$ electron per site). 
With blue crosses we denote points in the phase diagram for which we were able to obtain a converged sc-\dga \ solution.
Please note, that the sc-\dga \ method can be used both below the DMFT N\'eel temperature 
(indicated by the gray curve in~\cref{fig:phase-diagram}) 
as well as between the divergence lines (red and orange curves in~\cref{fig:phase-diagram}).
It is only on or directly next to divergence lines that we 
were not able to obtain convergence. 

The phase diagram in~\cref{fig:phase-diagram} serves as a proof of principle
and it is not our intention to discuss the sc-\dga \ results in the different parameter regimes
in the current paper. Instead, we show selected results for weak ($U=2$) and intermediate ($U=4$) coupling, 
where comparison to other methods is possible, as well as for strong coupling ($U=8$) 
and out of  half-filling ($n=0.85$) to show the applicability of the method in this 
interesting (e.g.\ with regard to   superconductivity) regime.

\subsection{Weak coupling}
In order to benchmark the method against known results, we first study a half-filled weak coupling case, with 
the interaction $U=2$ (in our units the bandwidth is $W=8$). 
This case was intensively studied by various methods in Ref.~\onlinecite{schaefer20}; 
and  in the spirit of Ref.~\onlinecite{schaefer20} we focus on 
spin fluctuations and the formation of the pseudogap at low temperature. 

In \cref{fig:square-u2-chi} the static magnetic susceptibility at 
${\bf q}=(\pi,\pi)$ is shown. For $U=2$  DMFT predicts a phase transition
at $T_N\approx 0.08$. The sc-\dga\ leads to a seemingly non-diverging antiferromagnetic (AFM) susceptibility;  
the updated self-energy in the BSE dampens the magnetic fluctuations and removes the divergence. 
In the temperature range accessible, the  sc-\dga\ susceptibility shows first a $1/(T-T_N)$ behavior, as in DMFT 
which has a finite N\'eel temperature $T_N$,
and then deviates to a linear behavior on the log-scale of \cref{fig:square-u2-chi}, 
corresponding to $\chi_m(T)\sim \exp(\alpha/T)$ with some constant $\alpha$. 
Such an exponential scaling with a divergence only at $T=0$ is to be expected 
for a two-dimensional system, fulfilling the Mermin-Wagner theorem \cite{Mermin1966}
(cf. also Fig. 13 in Ref.~\onlinecite{schaefer20}). 

The sc-\dga\ AFM susceptibility is somewhat smaller than the one from $\lambda$-corrected 
\dga\ presented in Ref.~\onlinecite{schaefer20} (not shown here) 
as well as slightly smaller than the parquet-\dga\ results 
(shown in \cref{fig:square-u2-chi} for 1 and  9 form factors). 
The overall behavior is however well reproduced. 

In order to correctly resolve the growing correlation length 
when lowering the temperature, the size of the momentum grid has to be increased. 
For the lowest two temperatures shown in \cref{fig:square-u2-chi} 
we performed extrapolation to infinite grid size (for details see Appendix~\ref{sec:appendixB}). 

With lowering the temperature the growing spin-fluctuations lead to enhanced scattering 
and suppression of the one-particle spectral weight at the Fermi energy 
and to opening of a pseudogap~\cite{Vilk1997,Rubtsov2009,Katanin2009,Rost2012,Schaefer2015-2,schaefer20}.
Due to the van Hove singularity~\cite{Gonzalez00,Halboth20b,Honerkamp01,Wu19} 
at the antinodal point ${\mathbf k}_\text{AN}=(\pi,0)$, 
the suppression happens earlier (upon lowering $T$) at this point 
than at the nodal point ${\mathbf k}_\text{N}=(\pi/2,\pi/2)$. In Fig.~\ref{fig:square-u2-se-scdga} we show the spectral functions (right) as well as the corresponding self-energies on the imaginary (Matsubara) frequency axis (left) for the two momenta ${\mathbf k}_\text{N}$ and ${\mathbf k}_\text{AN}$ and for different temperatures.

The pseudogap behavior of the spectral function is also visible 
in the imaginary part of self-energy on the Matsubara frequency axis. Upon lowering the temperature we first see metallic behaviour at both nodal and antinodal points: $|{\rm Im} \Sigma_N|$  at the first Matsubara frequency is smaller than at the second . At lower temperatures, the slope of ${\rm Im} \Sigma_N$ at the first two Matsubara frequencies changes sign; 
first only at the antinodal point (pseudogap) and finally at both nodal and antinodal points.
This is usually taken as a criterion for the opening of a pseudogap.

Note however that for $1/T=25$ there is already a pseudogap for ${\mathbf k}_\text{N}$ 
in Fig.~\ref{fig:square-u2-se-scdga} (top right) while the slope of ${\rm Im} \Sigma_N$
is still negative in Fig.~\ref{fig:square-u2-se-scdga} (top left). However,  a kink is visible. This kink of the
 analytic  $\Sigma_N$ function is apparently already enough for the analytic continuation to yield a  large negative  ${\rm Im} \Sigma_N$  at low real frequencies, which is needed for seeing a pseudogap.

In \cref{fig:square-u2-se} we show the behavior of the imaginary part 
of the self-energy at the first three Matsubara frequencies 
for the nodal and antinodal points as a function of inverse temperature. 
Here we compare the sc-\dga\ to parquet-\dga\ and $\lambda$-corrected
ladder-\dga~\cite{schaefer20}, and the diagrammatic Quantum Monte Carlo 
(DiagMC)~\cite{Simkovic_DiagMC,schaefer20}. For the first Matsubara frequency 
all the methods lie almost on top of each other down to approx. $1/T=10$ 
(at the nodal point differences  already become noticeable  at $1/T=10$). 
For lower temperatures the methods still qualitatively agree, 
but $|\mathrm{Im}\Sigma(\omega_n=\pi T)|$ grows faster in the $\lambda$-\dga\ 
and quantitatively agrees better with the DiagMC benchmark. 
In the sc-\dga, as well as in the p-\dga, this growth happens at lower temperatures. 
This is in correspondence to the behavior of AFM susceptibility, 
which also grows slower in these methods upon lowering the temperature 
compared to $\lambda$-\dga\ and DiagMC, while correctly reproducing the overall behavior.

If we however look at the two larger frequencies (middle and right panel of \cref{fig:square-u2-se}), 
the situation is opposite. Here both the p-\dga\ as well as sc-\dga\ follow the DiagMC benchmark closely 
up to $1/T=15$ and do not show any enhancement in $|\mathrm{Im}\Sigma|$ with lowering $T$, 
while in the  $\lambda$-\dga\ the 2$^{nd}$ and 3$^{rd}$ Matsubara frequency 
follow the behavior of the first one. This is probably a consequence of the $\lambda$-correction 
that is applied {\it a posteriori} to the self-energy. 
While it works very well for the AFM susceptibility and it gives the correct behavior 
for the low energy part in the self-energy, that is closely influenced (enhanced) by the strong 
spin fluctuations, it overestimates this influence for larger energies. 

All in all, the comparison with the DiagMC benchmark shows the ability of sc-\dga\ to describe the behavior of self-energy in all different temperature regimes: incoherent behavior at high temperatures, metalicity in the intermediate temperature regime and the opening of spin-fluctuation induced pseudogap. Quantitatively the agreement with DiagMC is excellent down to approx. $1/T=10$, with small quantitative differences visible for the lowest temperatures. The most pronounced difference is 
that the opening of the  pseudogap is shifted to lower temperature than in DiagMC. 

\begin{figure}
  \centering
  \includegraphics[width=0.48\textwidth]{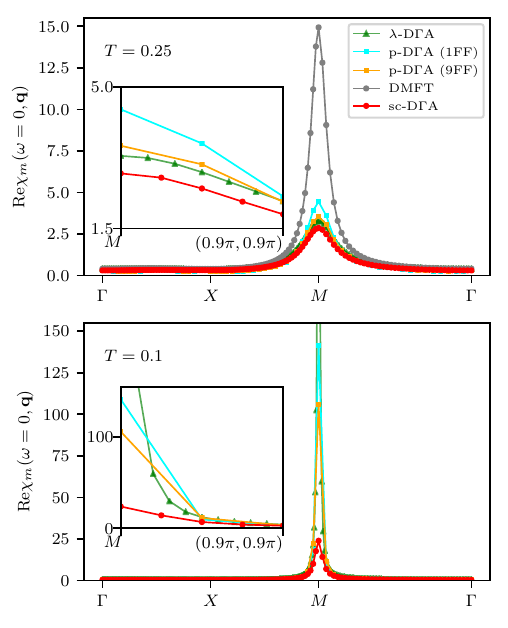}
  \caption{\label{fig:square-u4-chim-qpath} Static magnetic susceptibility
  for $U=4$
  on a path through the Brillouin zone for $T=0.25$ and $T=0.1$. The value of $\lambda$-\dga \ susceptibility at the $M$ point is $\chi_m(\omega=0, {\bf q} = (\pi,\pi))=415$ (beyond the y-range of the plot). A smaller momentum window is shown in the insets. }
\end{figure}

\begin{figure}
  \centering
  \includegraphics[width=0.48\textwidth]{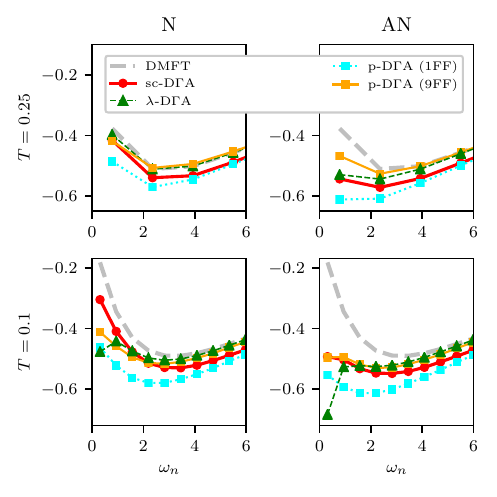}
  \caption{\label{fig:square-u4-se} Imaginary part of self-energy for the nodal (N) and antinodal (AN) points as a function of Matsubara frequencies for $U=4$, $n=1$ and two temperatures: $T=0.25$ and $T=0.1$. Different methods are distinguished by different symbols and colors. The 1FF and 9FF p-\dga \ data are reproduced from~Ref.~\cite{Eckhardt20}.}
\end{figure}

\subsection{Intermediate coupling}
Next, we increase the interaction to $U=4$ but stay at half-filling. Since we already
enter a regime, where the numerically exact methods are limited to high temperatures, we do not show comparisons to benchmarks. 
We focus here on the comparison to parquet-\dga\ and the $\lambda$-corrected
\dga. 

In \cref{fig:square-u4-chim-qpath} we show the static magnetic susceptibility as a function of momentum ${\bf q}$ for two temperatures. We choose $T=0.25$ for also comparing with the DMFT result that diverges for slightly lower temperature. Already for $T=0.25$ we see a large difference to the DMFT result. As for the different \dga \ methods, the results fall almost on top of each other with the exception of 1FF p-\dga, where the susceptibility is somewhat larger close to the $M$-point. For the lower temperature of $T=0.1$ the situation is quite different. Although all methods agree for momenta far from ${\bf q} =(\pi,\pi)$, close to it the results differ significantly, as it was the case for $U=2$. The sc-\dga \ susceptibility is again the smallest, followed by the p-\dga \ results.  

In  \cref{fig:square-u4-se} we show the imaginary part of self-energy as a function of Matsubara frequency for the same two temperatures as in \cref{fig:square-u4-chim-qpath}. For $T=0.25$ the \dga \ methods agree well, although not any more quantitatively as it was in the weak-coupling case for this temperature. Here the 1FF p-\dga \ result is noticeably different: at  $U=4$ the 1FF approximation is not sufficient any longer at this temperature (cf.~Ref.~\cite{Eckhardt20}). For $T=0.1$ at the antinodal point we already start to see the pseudogap behavior of self-energy in the sc-\dga \ and  p-\dga \ methods, whereas in $\lambda$-\dga\ the pseudogap sets in at a higher temperature of $T\approx 0.17$~\cite{Schaefer2016}. Except for the first Matsubara frequency, the three  \dga \ methods are in excellent, almost quantitative agreement. As in the $U=2$ case, the difference in the first Matsubara frequency is likely to be caused by much smaller AFM susceptibility in sc-\dga \ as compared to $\lambda$-\dga.           

An open question remains why the sc-\dga \ produces sizably smaller AFM susceptibility than the $\lambda$-\dga \ upon going to low temperatures. For the case of $U=2$ it is also significantly smaller than the DiagMC result~\cite{schaefer20}. An intuitive partial understanding can be gained by looking at the p-\dga \ results for one and nine form factors (1FF and 9FF). As already mentioned in Sec.IIIC and explained in~Ref.~\onlinecite{Eckhardt20}, for the 1FF approximation to  p-\dga \ the irreducible vertex $\Gamma$ is also local. But contrary to sc-\dga, it is updated after each update of the self-energy. Therefore when the damping effect of self-energy at low temperature becomes big, it can be counterbalanced by a larger $\Gamma$ which results in a larger susceptibility (cf. Figs.~\ref{fig:square-u2-chi} and~\ref{fig:square-u4-chim-qpath}). In sc-\dga \ this vertex stays the same throughout the calculation;  the two-particle feedback onto the self-energy is reduced~\footnote{In this case the Ward identity relating the irreducible vertex and self-energy $\Gamma=\frac{\delta\Sigma}{\delta G}$ is violated. This is also the case in the parquet approach, but certainly to a much smaller extent.}. There is also no feedback from the particle-particle channel that is present in p-\dga.
 
In the truncated unity p-\dga \ we can make $\Gamma$ systematically less local by using more form factors. It has also a strong effect on the susceptibility, as the 9FF p-\dga \ results show. In the case of $U=2$ the susceptibility is larger for 9FF, it is however smaller than the 1FF result for $U=4$ (cf. Fig.~\ref{fig:square-u4-chim-qpath}). Similar (opposite) tendencies of the AFM susceptibility were seen for the two values of $U$ in Ref~\onlinecite{Eckhardt20}. Although the convergence study in  Ref~\onlinecite{Eckhardt20} shows that at $T=0.25$ the 9FF p-\dga \ result is  converged with respect to the  number of form factors, it is quite likely not the case for much lower temperatures.   

In the $\lambda$-corrected \dga \ the vertex is also local and not updated. The imposed sum rule however imitates the mutual feedback of the one- and two-particle quantities.

\begin{figure}
  \centering
  \includegraphics[width=0.5\textwidth]{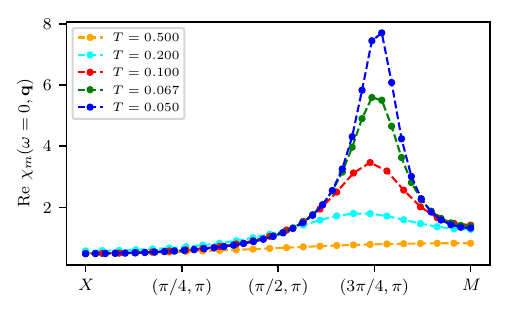}
  \caption{\label{fig:chi-qx-T} Static magnetic
    susceptibility in sc-\dga \ along the $X-M$ path in the Brillouin zone ($q_y = \pi$) for different temperatures $T$, $U=8t$ and $n=0.85$.}
\end{figure}
\begin{figure}
  \centering
  \includegraphics[width=0.5\textwidth]{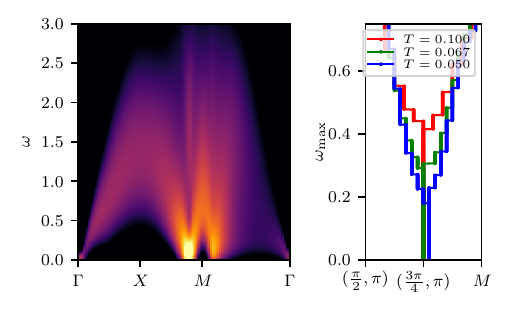}
  \caption{\label{fig:y-disp} Left panel: logarithmic plot of the 
    dynamic magnetic structure factor $\mathrm{Im} \chi_m(\mathbf{q}, \omega)/(1 - e^{-\omega/T})$
    at $U=8$, $n=0.85$ and $T=0.05$, obtained by analytic continuation.
    The analytic continuation was done with the maximum entropy method~\cite{Geffroy2019,kaufmannGithub}.
    Right panel: Y-shaped spin-excitation dispersion obtained from the  
    dynamic magnetic susceptibility at ${\bf q}=(q_x, \pi)$ for different temperatures $T$.}
\end{figure}
\begin{figure}[hb!]
  \centering
  \includegraphics[width=0.5\textwidth]{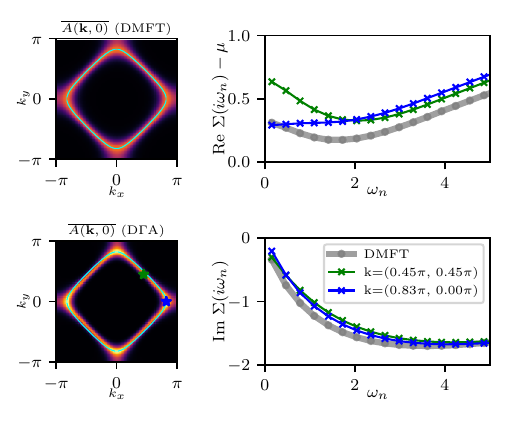}
  \caption{\label{fig:fs-se} Fermi surface (FS) and the corresponding 
    real and imaginary part of self-energy as a function of Matsubara frequency 
    as obtained in  DMFT and sc-\dga. 
    For sc-\dga \ we show two different momenta on the FS, 
    as indicated with green and blue stars on the FS plot. 
    The FS was obtained by plotting $\overline{A_{\bf k}(0)}\approx G({\bf k},\tau=1/(2T))$ (which avoids the analytical continuation and averages the spectral fucntion over an interval $\sim T$ around the FS).
    The non-interacting tight-binding FS is plotted with a thin cyan line 
    in both FS plots.  The parameters are $T=0.05$, $U=8$, $n=0.85$.}
\end{figure}
\subsection{Strong coupling}

Another interesting parameter regime that we can use the sc-\dga \ method for 
is the doped strong-coupling case,
which is relevant for superconductivity, as shown e.g.~in 
Refs.\ \cite{Gull2013, Chen2013, Otsuki2014, Jia2014, LeBlanc2019, Kitatani2019, Kitatani2020}.
Going to sufficiently low temperatures, such as in case of $\lambda$-\dga~\cite{Kitatani2019, Kitatani2020}, 
is a highly non-trivial task that requires computations with high numerical efficiency, 
since the momentum and frequency grids have to be sufficient to capture the growing correlation length. 

In the following we show results for the Hubbard model on a square lattice 
with $U=8$ and $15\%$ hole doping ($n=0.85$) in the temperature range 
$T\in [0.05,0.5]$. With lowering the temperature the magnetic fluctuations, 
still antiferromagnetic at $T=0.5$, become incommensurate. 
This is indicated by the shift of the maximum of the static magnetic susceptibility 
from $\bf{q}=(\pi,\pi)$ to ${\bf q}=(3\pi/4,\pi)$ in~\cref{fig:chi-qx-T}. 
If we look at the dynamic susceptibility $\chi_m(\omega, {\bf q})$ 
at finite frequencies $\omega$,  we can identify a splitting of the peak maximum. 
In the left panel of~\cref{fig:y-disp} we show the dynamic magnetic structure factor $\mathrm{Im} \chi_m(\mathbf{q}, \omega)/(1 - e^{-\omega/T})$, obtained by analytic continuation with the maximum entropy method~\cite{Geffroy2019,kaufmannGithub} for $T=0.05$ and also the position $\omega_{max}$ of the maximum 
(or maxima) as a function of ${\bf q}$ for different temperatures. 
The plots form characteristic $Y$-shaped spin-excitation dispersions, 
also seen experimentally~\cite{Chan2016} and discussed in Ref.~\cite{LeBlanc2019}. 
We observe that the frequency $\omega_{max}$, 
at which the splitting occurs, moves to lower values as the temperature is lowered. 
It could be interpreted as sharpening of the dispersion relation 
upon lowering the temperature.

In the right panels of \cref{fig:fs-se} the corresponding self-energy for the lowest temperature 
in~\cref{fig:chi-qx-T}, $T=0.05$, is shown. 
The imaginary part becomes slightly smaller at the lowest Matsubara frequencies in \dga.
In stark contrast to the particle-hole symmetric systems studied above,
the momentum dependence is rather small and visible mainly in the real part. 
This results in a slight deformation of the Fermi surface, which we can see
in the left panels of \cref{fig:fs-se}. While purely local correlations
cannot change the shape of the Fermi surface with respect to the tight-binding
model, non-local correlations of \dga\ in this case make the Fermi surface
slightly more ``quadratic'', since in the nodal direction the real part
of the self-energy at low frequencies is larger than DMFT. 
Furthermore, we observe that spectral weight is redistributed and more concentrated at the corners.

Our results demonstrate that sc-\dga\ works very well also in the doped case. This has been a weak spot for 1-\dga\ since in contrast to the symmetric half-filled model, non-local correlations change the filling. If the Coulomb interaction is rather large and we are close to half-filling, this effect is rather weak. Indeed previous 1-\dga\ calculations have hence focused on this parameter regime.  However, in other cases the filling of the DMFT serving as an input to the one-shot calculation can and will be quite different from the filling of the  1-\dga.  This renders a self-consistent treatment with an adjustment of the chemical potential obvious, so that the filling remains as that for which the vertex $\Gamma$ was calculated.

\section{Multi-orbital calculations}
\label{sec:results-multi}
\subsection{Two-orbital model}

\begin{figure}
  \centering
  \includegraphics[width=0.5\textwidth]{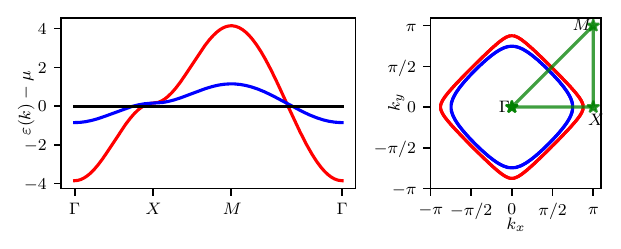}
  \caption{\label{fig:2band-tb} Two-band model: band structure (left) and 
  Fermi surfaces at $T=0.1$ (right). The filling is $n=1.7$, 
  such that the chemical potential is slightly temperature-dependent.}
\end{figure}
\begin{figure}
  \centering
  \includegraphics[width=0.45\textwidth]{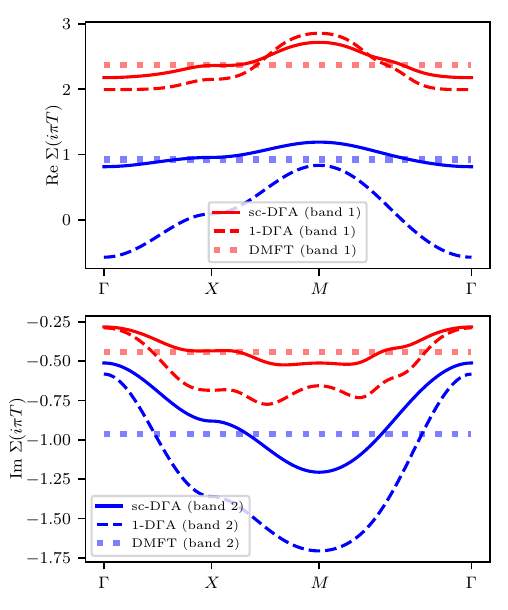}
  \caption{\label{fig:2band-siw-b10} Real (top) and imaginary part (bottom)
    of the self energy at the lowest Matsubara frequency 
    for the two band Hubbard model at $T=0.1$ along a high-symmetry path through the Brillouin zone.}
\end{figure}
\begin{figure*}
  \centering
  \includegraphics[width=0.8\textwidth]{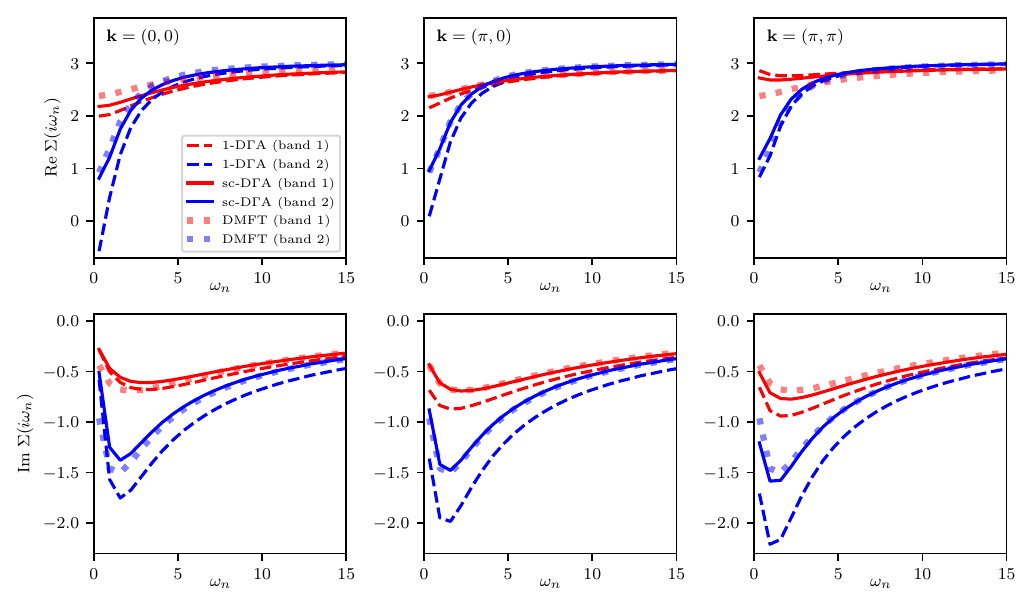}
  \caption{\label{fig:2band-siw-gamma} Real (top) and imaginary (bottom) part of the 
    self-energy vs.\ Matsubara frequency at $T=0.1$ for the 
    $\Gamma$, $X$ and $M$-point, comparing DMFT,  \osdga\, and  \scdga for a two-band Hubbard model. 
    The different colors refer to the two bands, 
    as plotted in \cref{fig:2band-tb}; the different line types to the different methods.}
\end{figure*}

In order to demonstrate that self-consistent D$\Gamma$A also works for
more than one orbital, we consider next a
simple two-orbital model on a square
lattice. 
Here, electrons can hop only to neighboring atoms
with hopping amplitudes  $t_1=1$ and $t_2=0.25$ for the two orbitals.
This gives rise to a wide and a narrow cosine band
with band-width 8 and 2, respectively.
Along a high-symmetry path, the bandstructure is shown in \cref{fig:2band-tb} (left) and the Fermi surface of the non-interacting tight-binding model in \cref{fig:2band-tb} (right). This tight-binding model is supplemented by a 
Coulomb repulsion
parametrized in the Kanamori form with intra-orbital
interaction $U=4$, Hund's coupling $J=1$, and inter-orbital interaction $V=U-2J$. 
The spin flip and pair hopping processes are of the same magnitude $J$.
Considering the different band widths,  the wide band will be weakly correlated,
since $U$ is only one half of the band width. The narrow band,
however, is strongly correlated since $U$ is twice as large as its band width.

\begin{figure*}
  \includegraphics[width=0.9\textwidth]{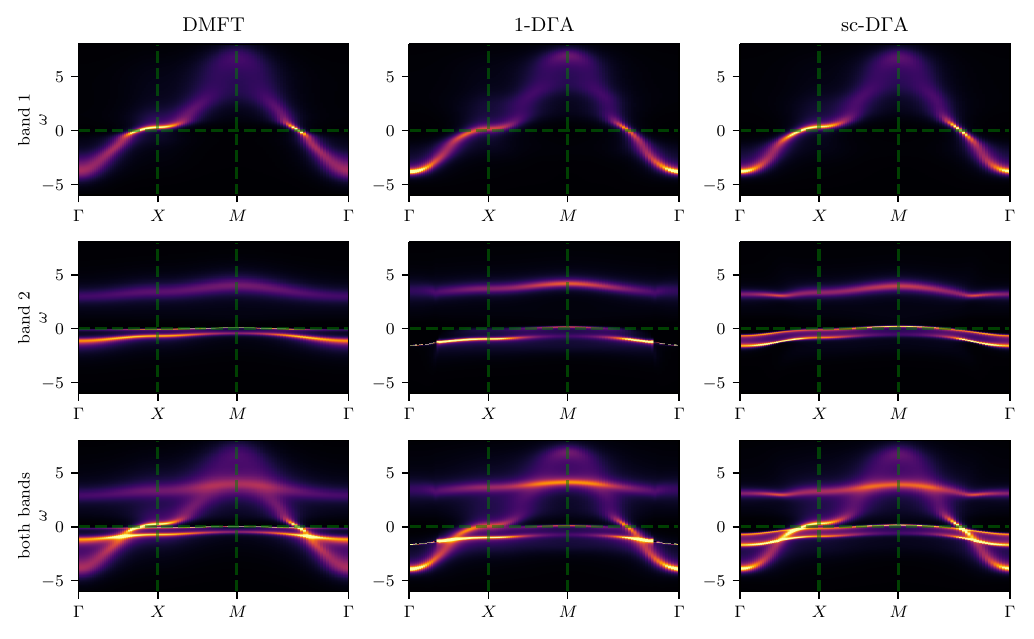}
  \vspace{-0.2cm}
  \caption{\label{fig:2band-full} Momentum-resolved spectral function of the two-orbital Hubbard model
  at $T=0.1$ along a high-symmetry path through the Brillouin zone.
  The first column shows only the component corresponding to the wide band,
  the second column only the component corresponding to the narrow band. 
  In the third row the full spectral function is shown.
  }
\end{figure*} 
\begin{figure*}
  \includegraphics[width=0.9\textwidth]{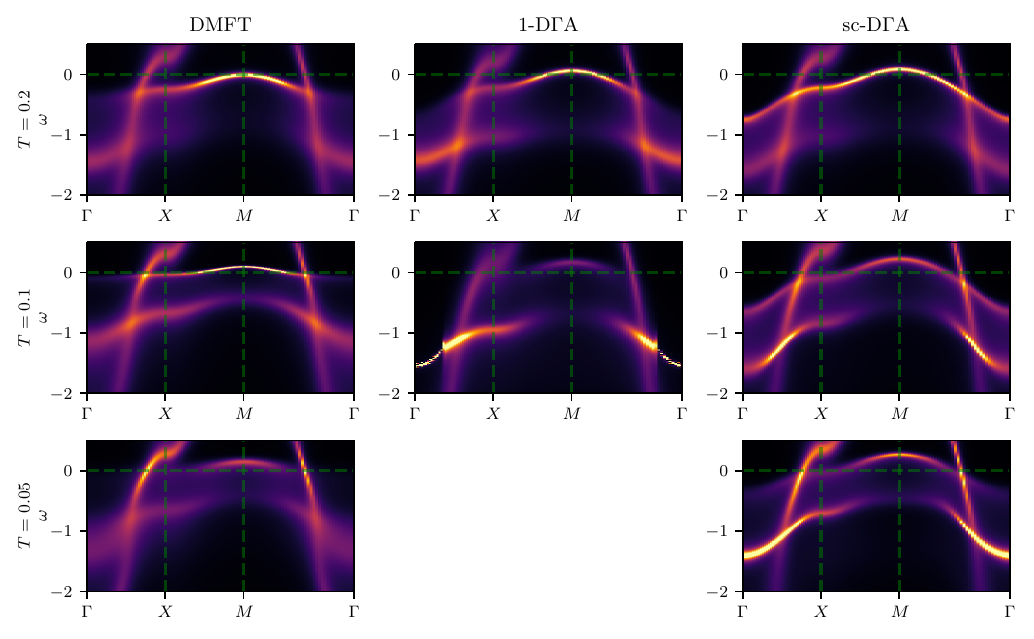}
  \vspace{-0.2cm}
  \caption{\label{fig:2band-zoom} Low-frequency zoom of momentum-resolved spectral function for the two-orbital Hubbard model
  at
  temperatures $T=0.2, 0.1, 0.05$ on a path through the Brillouin zone. In the \osdga-spectrum at $T=0.1$ it is clearly visible
	that the analytic continuation does not work well in the vicinity of the $\Gamma$-point. 
	This is a typical issue for \osdga\ calculations where non-local self-energy corrections become large.}
\end{figure*}

In the context of an orbital-selective Mott transition 
\cite{Anisimov2002,Liebsch2003a,Liebsch2005,Koga2005,Biermann2005,Arita2005,Knecht2005,Sakai2006,Ferrero2005,Costi2007,Koga2004,Greger2013,Valli2015,Tocchio2016,Philipp2017,Hu2017},
such simple half-filled two-band models with different bandwidths and intra-orbital hopping have been studied very intensively in DMFT.
Early calculations however did not include the spin flip and pair hopping processes, 
but only the density-density interactions for technical reasons. 
In this situation, the tendency toward an orbital selective Mott transition 
is largely exaggerated: a spin $S_z=\pm1$ formed by the Hund's exchange 
cannot undergo a joint SU(4) Kondo effect, while the spin-1 of the 
SU(2)-symmetric interaction can. 
As we are primarily interested in testing the \scdga\ method, we consider here the case
where the model is doped away from half-filling or $n=2$ electrons per site. Specifically, we consider the doping $n=1.7$. 
This gives rise also to a non-zero real part of the self-energy and (slightly) different fillings of the two orbitals; and hence tests various aspects at the same time.

In \cref{fig:2band-siw-gamma} we show the self-energy at selective $\mathbf k$-points. 
For the given parameters, the  \osdga\ corrections to the self-energy are extremely strong,
even exceeding the value of the DMFT self-energy. 
The reason for this is that we are quite close to an (incommensurate)
antiferromagnetic phase transition in DMFT. 
Immediately before the phase transition, 
the \osdga\ corrections become even larger and turn the system insulating.

Similar as for the one-band model, the self-consistency suppresses 
the antiferromagnetic fluctuations; the actual phase transition occurs 
only at zero temperature because we are in two dimensions. 
Hence the \scdga\ corrections are much weaker at the fixed temperature 
close to the DMFT phase transition. They will, as a matter of course, 
become stronger at lower temperatures which are not reachable by
\osdga\ exactly because of the DMFT phase transition. 
Indeed, \cref{fig:2band-siw-gamma} suggests that \scdga\ is not too 
distinct from the DMFT result. That is, the self-consistency dampens away much of the one-shot corrections.

However, there is actually a quite important difference: 
Depending on the $\mathbf k$-point the \scdga\ imaginary part of 
the self-energy at low Matsubara frequencies is above or below the 
DMFT self-energy in  \cref{fig:2band-siw-gamma}. 
This becomes even more obvious in \cref{fig:2band-siw-b10}, 
where we plot  the self-energy at the lowest  Matsubara frequency and 
see that the low frequency self-energy strongly depends on the momentum.
A strong momentum differentiation of the imaginary part of the 
self-energy (i.\ e.\ the scattering rate) has also been reported for a SrVO$_3$ monolayer~\cite{Pickem2020arxiv}.

In contrast to the imaginary part, the real part of the self-energy only shows a weak 
momentum dependence around the DMFT value in  \cref{fig:2band-siw-b10}.
This is different for \osdga\ where the strong corrections are
also reflected in a sizable momentum-dependence of the real 
part of the self-energy; the strongly correlated band (band 2; blue) 
also displays a sizable overall shift compared to the DMFT result in \osdga. 

But let us turn back to the momentum dependence of the self-energy in \scdga. It has a larger influence 
on the spectral function (\cref{fig:2band-full}) than what one might expect
from the Matsubara-frequency dependence in \cref{fig:2band-siw-gamma}.
In \cref{fig:2band-full} we see, for all three methods,
that the weakly correlated band 1 is still close to the tight-binding 
starting point in \cref{fig:2band-tb}, whereas the strongly correlated 
band 2 is split into an upper Hubbard band (around $\omega\sim 4$), 
a lower Hubbard band (around $\omega=-0.5$), 
and a central quasiparticle peak around the Fermi level ($\omega=0$).
The last is better visible in the zoom-in provided by \cref{fig:2band-zoom}.
The aforementioned momentum differentiation of the self-energy results in a 
considerably wider central quasiparticle band in \scdga\ than in DMFT or \osdga. 
In \osdga\, the strong fluctuations around the phase transition 
also smear out the central band when reducing temperature from 
$T=0.2$ to $T=0.1$;
$T=0.05$ is below the DMFT ordering temperature and a one-shot calculation 
is hence no-longer possible (the reduction of the N\'eel temperature and susceptibility 
requires the self-consistency or a Moriya $\lambda$-correction \cite{Katanin2009}).

\begin{figure}
  \centering
  \includegraphics[width=0.5\textwidth]{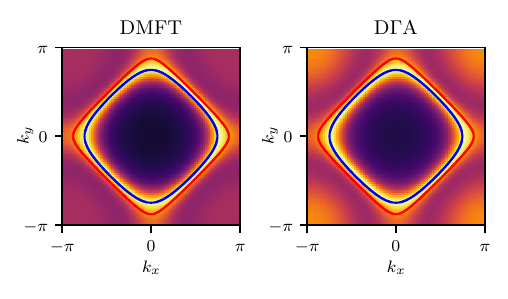}\\
  \includegraphics[width=0.5\textwidth]{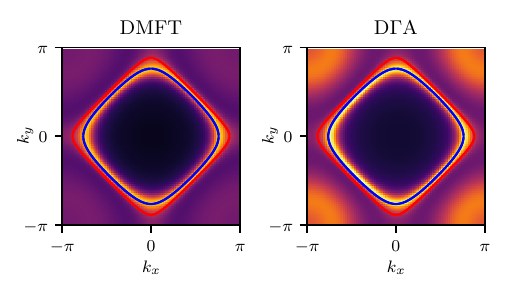}
  \caption{\label{fig:fs-2band} Spectral weight at the Fermi level 
    $\overline{A({\mathbf k},0)}\approx G({\mathbf k},\tau=1/2T)$
    of the two-band Hubbard model for $T=0.1$ (top) and $T=0.05$ (bottom),
    comparing DMFT (left) and \scdga\ (right) and the tight-binding model without interaction (red and blue lines).}
\end{figure}

In \cref{fig:fs-2band}, we further show the spectral weight at the Fermi level 
in DMFT and \scdga, summed over both orbitals.
Clearly a Fermi surface close to the tight binding ones is visible. 
This stems mostly from the wide, less correlated band. 
The narrow, strongly correlated band is slightly shifted downwards
to lower energy and considerably broadened, cf.~\cref{fig:2band-zoom}.
Since the band is so flat, this tiny shift results in a sizeable deformation
of the spectral weight distribution on the Fermi level:
Considering also that
$\overline{A({\mathbf k},0)}$ averages over a frequency range $\sim T$,
we get diffuse arcs around the $M$-point, i.e., $(\pi,\pi)$, 
which is visible in \cref{fig:fs-2band}.
However, due to the strong renormalization that is already present in DMFT,
the narrow band gives only a small contribution to the spectral weight
on the Fermi level. 

\begin{figure}
  \centering
  \includegraphics[width=0.5\textwidth]{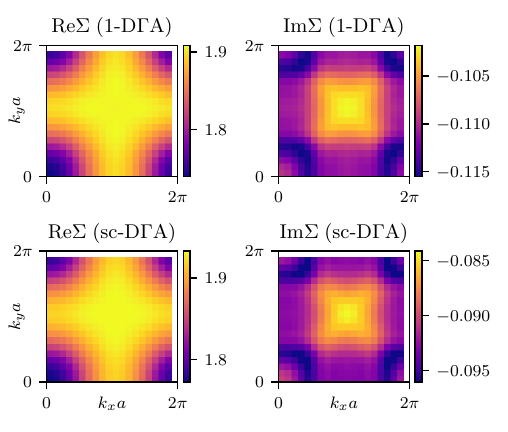}
  \caption{\label{fig:se-svo} Momentum dependence of the self-energy
  of strontium vanadate at the lowest positive Matsubara frequency and $k_z=0$. 
  Upper row: one-shot abinitioD$\Gamma$A,
  lower row: self-consistent D$\Gamma$A. The DMFT value is (1.861-0.104$i$)eV.
  }
\end{figure}

\begin{figure*}
  \centering
  \includegraphics[width=\textwidth]{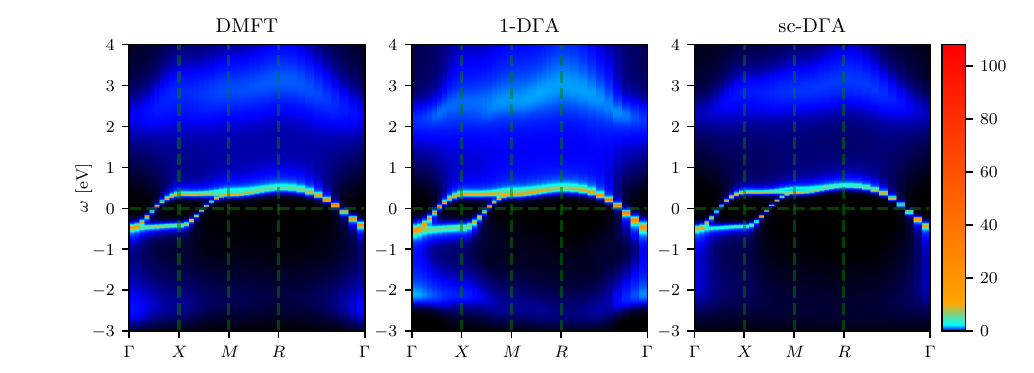}
  \caption{\label{fig:akw-svo} Momentum-dependent spectral function of SrVO$_3$,
  comparing DMFT, one-shot \dga, and self-consistent \dga.}
\end{figure*}

\subsection{Strontium vanadate}
As a second, archetypical multi-orbital application we study
bulk strontium vanadate SrVO$_3$ at room temperature ($T=26.3$meV).
This material has served as a testbed for the development 
of realistic materials calculations with strong correlations, 
and is hence most intensively studied 
\cite{Sekiyama04,Pavarini04,Nekrasov05a,Maiti06,Leonov11,Karolak11,
Lee12,Casula12,Tomczak12,Taranto13,Miyake13,Sakuma13,Tomczak14,
Wadatti14,Ribic14,Zhong15,Kazuma16,Boehnke16,Nakamura16,Backes16,
Bhandary2016,Tomczak17,Kaufmann2017,Bauernfeind17,Sim2019}. 
Also the first realistic materials calculations using diagrammatic 
extensions of DMFT, i.e., {\em ab initio} D$\Gamma$A, 
have been performed for this perovskite \cite{galler17}. 
SrVO$_3$ is a strongly correlated metal with a quasiparticle 
renormalization of about two \cite{Sekiyama04}. 
Electronic correlations also lead to a kink in the self-energy 
and energy-momentum dispersion relation \cite{Nekrasov05a,Byczuk2007,Aizaki12,Held13}. 
Theoretical calculations and experiments do not indicate any long-range order.

For this realistic {\em ab initio} calculation, we start with a Wien2K calculation  \cite{blaha2001wien2k,schwarz02}  using the
PBE exchange correlation potential in the generalized gradient 
approximation (GGA) \cite{Perdew96}, and a lattice constant of $a=3.8$\AA.
The calculated bandstructure is  projected onto maximally localized $t_{2g}$ Wannier orbitals
\cite{Pizzi2020,Marzari12,mostofi2008wannier90} using wien2wannier \cite{kunevs2010wien2wannier} 
This three-band Wannier Hamiltonian, 
available open source \footnote{See \url{https://github.com/AbinitioDGA/ADGA/blob/master/srvo3-testdata/srvo3_k20.hk}},
is supplemented by a 
Kanamori Coulomb interaction including the same terms as for the 
two-band model and parameterized by $U'=3.5$eV, $J=0.75$eV and corresponding $U=U'+2J=5$eV. The interactions $U'$ and $J$ have been calculated by the constrained local density approximation (cLDA) in \onlinecite{Sekiyama04}; $J$ was later slightly corrected as outlined in Section~4.1.3 of \cite{held2007electronic} to account for the precise way $J$ enters in Hamiltonian (\ref{eq:hubbard-model}) and the cLDA.
The difference to earlier  \abinitiodga\ \cite{galler17,Galler18,galler19}
calculations, which have been one-shot non-self-consistent calculations,
is that we now perform a self-consistent calculation.

As already mentioned, a Moriya-$\lambda$ correction is extremely difficult
for such realistic multi-orbital calculations. 
There is not only a magnetic and charge $\lambda$ for every orbital 
but additionally also various orbital combinations. 
Hence, we hold that a self-consistent calculation shall be preferable 
compared to a high-dimensional fit of the various $\lambda$ parameters. 
Also conceptionally it is a clearer approach.

In \cref{fig:se-svo} we compare the self-energy of the one-shot and self-consistent
\dga\ calculation. In contrast to the two-band Hubbard model study above, 
the differences are here only minor. The reason for this is that in case of the 
two-band Hubbard model we were close to the DMFT phase transition, 
whereas SrVO$_3$ is rather far away from any phase transition. 
Hence, the \osdga\ corrections are much smaller to start with. 
In such a situation, the self-consistency is not necessary. 
This justifies {\em a posteriori} the use of non-self-consistent
\dga\ in Refs.~\cite{galler17,Galler18,galler19}.

Nevertheless, \cref{fig:akw-svo}
indicates some minor differences between the DMFT,  \osdga\, and  \scdga\ spectral functions. There are minor differences between  \osdga\ and  \scdga\ regarding the 
weight of the lower Hubbard band and the broadening of the
quasiparticle peak. 
This behavior is perfectly in-line with the effect of \osdga\ in other systems
studied above.
Furthermore there is a shift of the position of the lower Hubbard band toward lower binding energies visible at the $\Gamma$ point.
Experimentally, the maximum of the lower Hubbard band is slightly above -2$\,$eV
\cite{Sekiyama04}.

\section{Conclusion}
We have presented a self-consistent solution of the
ladder \dga\ equations where the calculated  \dga\ self-energy  is fed back into the Bethe-Salpeter ladder. This dampens the Green's function and thus the overall strength of the ladder, largely reducing the critical temperatures of DMFT. Hitherto, a similar effect has been achieved by a Moriyaesque $\lambda$ correction for one band-models; multi-orbital models have only been studied by  one-shot, non-self-consistent and non-$\lambda$-corrected calculations. Applying such a   $\lambda$ correction to multi-orbital or doped systems is difficult, to say the least. One-shot calculations, on the other hand, are disputable whenever the 
non-local corrections to DMFT become large.
Our paper demonstrates that conceptionally clean self-consistent calculations are  indeed feasible and work well, also for  multi-orbital and doped systems.

For the one-band Hubbard model we have benchmarked the method against previous  ($\lambda$-corrected and parquet)  \dga\ and numerically exact DiagMC results at weak coupling. We find an excellent agreement up to the point where the susceptibilities become huge, where
self-consistent \dga\ yields a somewhat reduced susceptibility. The self-consistency allows applying \dga\ even in the close vicinity of the divergence lines of the vertex, at strong coupling and for doped systems.

For the two-band Hubbard model we study the regime close to the DMFT phase transition. Here, the one-shot  \dga\ corrections are large but the self-consistency mitigates this to a large extent. While the frequency dependence eventually looks similar to that of DMFT, there is a sizable momentum dependence which leads to a widening of the quasiparticle band. In case of SrVO$_3$ we have performed realistic \abinitiodga\ materials calculations. Here, we are not close to any phase transition and the difference between one-shot and self-consistent \abinitiodga\ is minute.

\acknowledgements
We thank Thomas Sch\"afer, Fedor \v{S}imkovic and Patrick Chalupa for providing
reference data; and Patrik Thunstr\"om, Anna Galler  and Jan M.\ Tomczak for fruitful discussions and valuable advice.
JK further thanks Oleg Janson, Alexander Lichtenstein, Andrey Katanin, Jan Kune\v{s}, Evgeny Stepanov, Patrik Gunacker, Tin Ribic, Dominique Geffroy and Benedikt Hartl
for fruitful discussions.
This work has been supported  financially by the European Research Council under the European Union's Seventh Framework Program (FP/2007-2013)/ERC grant agreement n.\ 306447 and by the 
Austrian Science Fund (FWF) through 
projects P 30997, P 30819 and P 32044. 
Calculations have been done in part on the Vienna Scientific Cluster (VSC).

\appendix
\section{Implementation}
\label{sec:implementation}
For the practical evaluation of the \dga\ equations
\crefrange{eq:bseq-ladder-dga}{eq:schwidy},
we use the \abinitiodga\ code \cite{galler19}. 
Here we describe the details
of the implementation, which are closely connected
to \abinitiodga.
Solving the aforementioned equations self-consistently
means that the \abinitiodga\ code is executed several times
in a loop in order to do a fixed-point iteration.
Before each iteration, we create an updated trial input,
until the point where the output does not differ from the input any more.
Therefore, in order to describe the details of the updates,
we have to recapitulate the input structure of \abinitiodga\ first.

Apart from the system-defining parameters (tight-binding Hamiltonian
and U-matrix) the following quantities are required as input:
\begin{enumerate}   
  \item lattice self-energy $\Sigma^k$ 
    (can also be momentum-independent)
  \item impurity self-energy $\Sigma_\text{imp}^\nu$
    (can be identical to the lattice self-energy, as in 1-\dga)
  \item impurity Green's function $g^\nu$
  \item impurity two-particle Green's function $G^{\nu\nu'\omega}$
\end{enumerate}
The update proceeds in the two steps described in the following.

\subsection{Update of the self-energy and one-particle Green's function}
This step defines the update. We take trial and result self-energies from
several preceeding iterations and compose a new trial self-energy
$\Sigma_\text{trial}^{k\,(j)}$ for the $j$-th iteration.
This is prediction is usually made by the Anderson acceleration algorithm
\cite{Anderson1965,Walker2011}
(also known as Pulay-mixing \cite{Pulay1980} or direct inversion in iterative subspace, DIIS \cite{Csaszar1984}).
This trial self-energy is then used to compute a new local propagator 
$G_\text{loc}^{\nu\, (j)}$ by
\begin{equation}
  \label{eq:update-gloc}
  G_\text{loc}^{\nu\, (j)} = \frac{1}{V_\text{BZ}} \int \dif^d{k}
    \Big[ (i\nu + \mu^{(j)})\mathbbm{1} - h(\mathbf{k})
    - \Sigma_\text{trial}^{k\, (j)} \Big]^{-1},
\end{equation}
where the chemical potential $\mu^{(j)}$ is adapted such that the
expectation value of the particle number stays at the desired value.
The change of the chemical potential usually stays in the range of a few percent.
Once the new local Green's function is determined, we project (downfold) it
to the correlated impurity subspaces. Thus, each impurity $I$ obtains
its new Green's function $g_I^{\nu\,(j)}$.

\subsection{Update of impurity quantities}
This step is inherent to our specific implementation of \abinitiodga,
and not part of the algorithm \emph{per se}. But since \abinitiodga\ reads
the one- and two-particle Green's function instead of the irreducible vertex,
we need to ``wrap'' the irreducible vertex (unchanged throughout all iterations) 
in the new impurity propagator by means of the Bethe-Salpeter equation. 
In order to avoid direct computation
of the irreducible vertex, we compute the updated generalized susceptibility 
for iteration $(j)$ in channel $r$ by
\begin{equation}
  \label{eq:update-imp-susc}
  \chi_r^{(j)} = \chi_r^\text{DMFT}
    \Big[ \chi_r^\text{DMFT} + \chi_0^{(j)}
      - \chi_0^{(j)} \big(\chi_0^\text{DMFT}\big)^{-1} \chi_r^\text{DMFT} \Big]^{-1}
    \chi_0^{(j)}.
\end{equation}
Note that all susceptibilities in this equations are compound-index matrices
in the orbital space of the impurity and fermionic frequencies.
The new impurity one-particle Green's function enters into this equation
only through $\chi_0^{(j)}$ of \cref{eq:chi0}, where updated impurity
Green's functions $g^{(j)}$ are used.
The two-particle Green's function is obtained by dividing through $\beta$
and adding a disconnected part, according to \cref{eq:gen-susc}.

Furthermore, it is necessary to compute an updated (``fake'') impurity self-energy
by the equation of motion. The reason for this can be seen in Eq.\ (75) of
Ref.\ \cite{galler17}. There, the DMFT self-energy appears as a separate
term. However, in its essence it is not the DMFT self-energy, but rather
the result of the Schwinger-Dyson equation of motion for the impurity\footnote{
  We thank Patrik Thunstr{\"o}m for drawing our attention to this crucial insight.}.
In Ref.\ \cite{galler17}, this term is subtracted and substituted by the
actual DMFT self-energy, in order to mitigate effects of finite frequency boxes.
Therefore, we compute the impurity self-energy from the equation of motion,
\begin{equation}
  \label{eq:eom-impurity}
  \Sigma_{\text{con},m,I}^{\nu\,(j)} = \frac{1}{\beta} \sum_{\nu'\omega} \sum_{lhn}
    U_{I,mlhn} G_{\text{con},I,nlhm}^{\nu'\nu\omega\,(j)} / g_{I,m}^{\nu\,(j)}
\end{equation}
using both the new $(j)$ and the DMFT one- and two-particle Green's function. 
The index $I$ labels the $I$-th impurity of the unit cell.
Importantly,
the frequency boxes have to be identical. Then the difference of these two
self-energies is added to the DMFT self-energy and taken as the new (fake) impurity
self-energy. In this way the effects of finite-box summation are cancelled.
We emphasize that the ``fake'' impurity self-energy is merely an auxiliary quantity and
never used to extract any physical properties of the result. Only the lattice
self-energy is subject to physical interpretation in our computations.

\section{Extrapolation of the susceptibility}
\label{sec:appendixB}
\begin{figure}[!h]
  \includegraphics[width=0.5\textwidth]{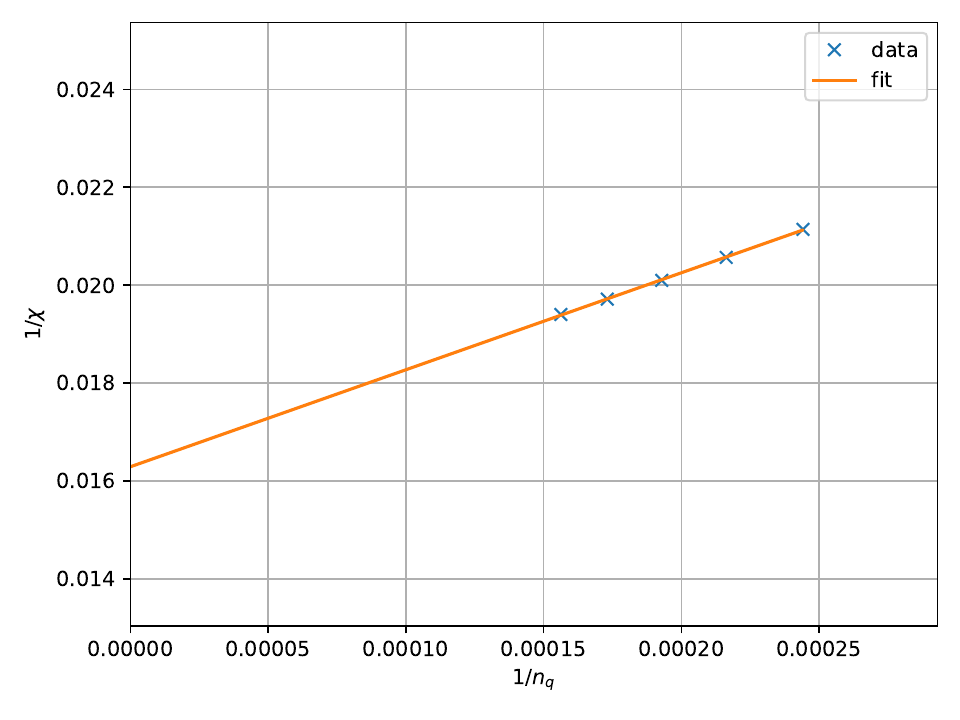}
  \caption{\label{fig:chi-extrapolation-b20} Extrapolation of antiferromagnetic susceptibility of the
    square-lattice Hubbard model with $U=2$ at $T=0.05$.
    The largest $\mathbf{q}$-grid is 80$\times$80.}
\end{figure}

\begin{figure}[!h]
  \includegraphics[width=0.5\textwidth]{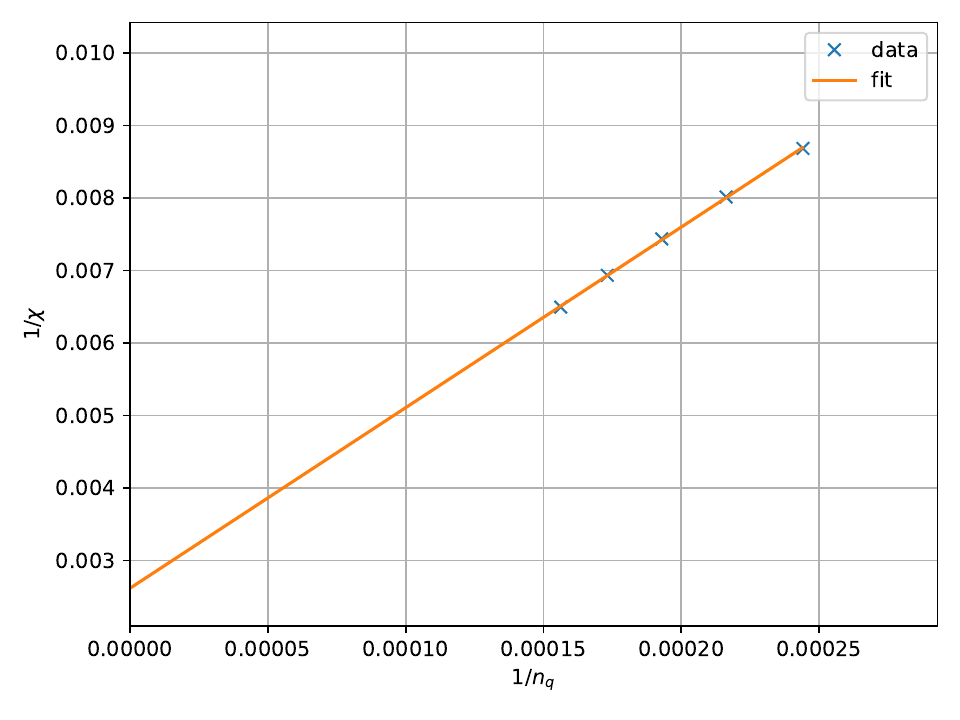}
  \caption{\label{fig:chi-extrapolation-b25} Extrapolation of antiferromagnetic susceptibility 
    of the square-lattice Hubbard model with $U=2$ at $T=0.04$.
    The largest $\mathbf{q}$-grid is 80$\times$80.}
\end{figure}

Since we are quite limited in the number of $\mathbf{q}$-points 
that we can use in our calculation, we have to do an
extrapolation of the magnetic susceptibility. This
is possible due to the observation that the inverse
of the antiferromagnetic susceptibility depends linearly on the inverse of the
number of $\mathbf{q}$-points. 
In particluar, the extrapolation was necessary for sc-\dga\ on the square-lattice
Hubbard model with $U=2$ at $T=0.05$ and $T=0.04$.
There the \dga\ calculation was done with 48$\times$48, 
64$\times$64, 68$\times$68, 72$\times$72, 76$\times$76, 80$\times$80
$\mathbf{k}$- and $\mathbf{q}$-points. In \cref{fig:chi-extrapolation-b20}
and \cref{fig:chi-extrapolation-b25} it is visible
that the extrapolation with above mentioned linear relation is indeed possible.
Although a deviation from this behavior is to be expected as $n_\mathbf{q}\rightarrow\infty$,
it can only lead to a small change in the logarithmic plot in~\cref{fig:square-u2-chi} and thus our conclusions
remain unchanged.

On the other hand, 
for $\mathbf{k}$- and $\mathbf{q}$-grids of 48$\times$48 or larger,
we find that the self-energy is
practically independent on the number of $\mathbf{k}$- and $\mathbf{q}$-points, such 
that no extrapolation is necessary there. 
\FloatBarrier
\bibliography{paper}{}
\end{document}